\documentclass[%
 reprint,
superscriptaddress,
nofootinbib,
 amsmath,amssymb,
 aps,
pra,
floatfix,
]{revtex4-2}

\usepackage{graphicx}
\graphicspath{ {./figures/} }

\usepackage{dcolumn}
\usepackage{bm}
\usepackage[utf8x]{inputenc}
\usepackage{amsmath}
\usepackage{graphicx}
\usepackage{mathtools} 
\usepackage{color}
\usepackage{bm}
\usepackage{braket}
\usepackage{float}
\usepackage{bigints}
\usepackage{ragged2e}
\usepackage{textcomp,gensymb}
\usepackage{appendix}
\usepackage{upgreek}
\usepackage{rotating}
\usepackage{amsmath}
\usepackage{subfig}
\usepackage[backgroundcolor=orange]{todonotes}
\usepackage{siunitx}
\usepackage{comment}
\usepackage{footmisc} 
\usepackage{chngcntr}

\usepackage{hyperref}

\begin{document}


\title{Minimising interference in low-pressure supersonic beam sources}

\author{\underline{Jack Kelsall} \textsuperscript{$\ast,$}}
\affiliation{Cavendish Laboratory, Department of Physics, University of Cambridge, JJ Thomson Ave, Cambridge, UK}

\author{\underline{Aleksandar Radi\'{c}} \textsuperscript{$\ast\,,\dagger$}}
\affiliation{Cavendish Laboratory, Department of Physics, University of Cambridge, JJ Thomson Ave, Cambridge, UK}

\author{John Ellis}

\author{David J. Ward}

\author{Andrew P. Jardine}
\affiliation{Cavendish Laboratory, Department of Physics, University of Cambridge, JJ Thomson Ave, Cambridge, UK}
\footnotetext{\noindent\textsuperscript{$\ast$}These authors contributed equally to this work.}
\footnotetext{\noindent\textsuperscript{$\dagger$}Corresponding author: ar2071@cam.ac.uk}

\date{\today}
\begin{abstract}
\noindent
Free-jet atomic, cluster and molecular sources are typically used to produce beams of low-energy, neutral particles and find application in a wide array of technologies, from neutral atom microscopes to instruments for surface processing. We present a simple analytical theory that is applicable to many of these sources, when (i) the nozzle-skimmer distance is such that free molecular flow is achieved and (ii) there is negligible interference within the skimmer itself. The utility of the model is demonstrated by comparing experimental data with calculations performed using the theory. In particular, we show that skimmer interference is negligible compared to attenuation by `background' gas for room-temperature beams. Our treatment does not depend on any free parameters and obviates the complexity of previous theories. As a result, we are able to devise a number of design recommendations to minimize interference in sources operating with cryogenic-temperature beams.
\end{abstract}

\maketitle
	
\section{Introduction}

A free-jet, or supersonic, source generates a thermal beam of neutral particles from a reservoir of gas and is used in a range of atomic, molecular and ion beam instruments. Free-jet sources have become the dominant mechanism for neutral species beam generation, with mostly empirical developments driving novel experimental techniques such as atom lithography \cite{anab,Allred2010,Berggren_1995}, helium atom micro-diffraction \cite{von_jeinsen_2d_2023,defects_radic,2d_methods,FLATABO2024113961}, scanning helium microscopy \cite{Palau2023,radic_3d_2024,LambrickDiffuse2022,FLATABO2024113961} and ultrahigh resolution spectroscopy \cite{coldmolbeams,bismuthhas,liu_distinguishing_2024,liu_experimental_2024}. A thermal beam consists of particles with a low energy, given by $E\sim k_BT$, which gives beam energies of $\approx\SI{60}{\milli\electronvolt}$ for \textsuperscript{3}He from a room-temperature reservoir \cite{von_jeinsen_2d_2023}. The probe particles are therefore unable to penetrate into any sample, leading to a non-destructive and exclusively surface-sensitive probe.

\begin{figure}[h]
	\centering
	\includegraphics[width=0.495\textwidth]{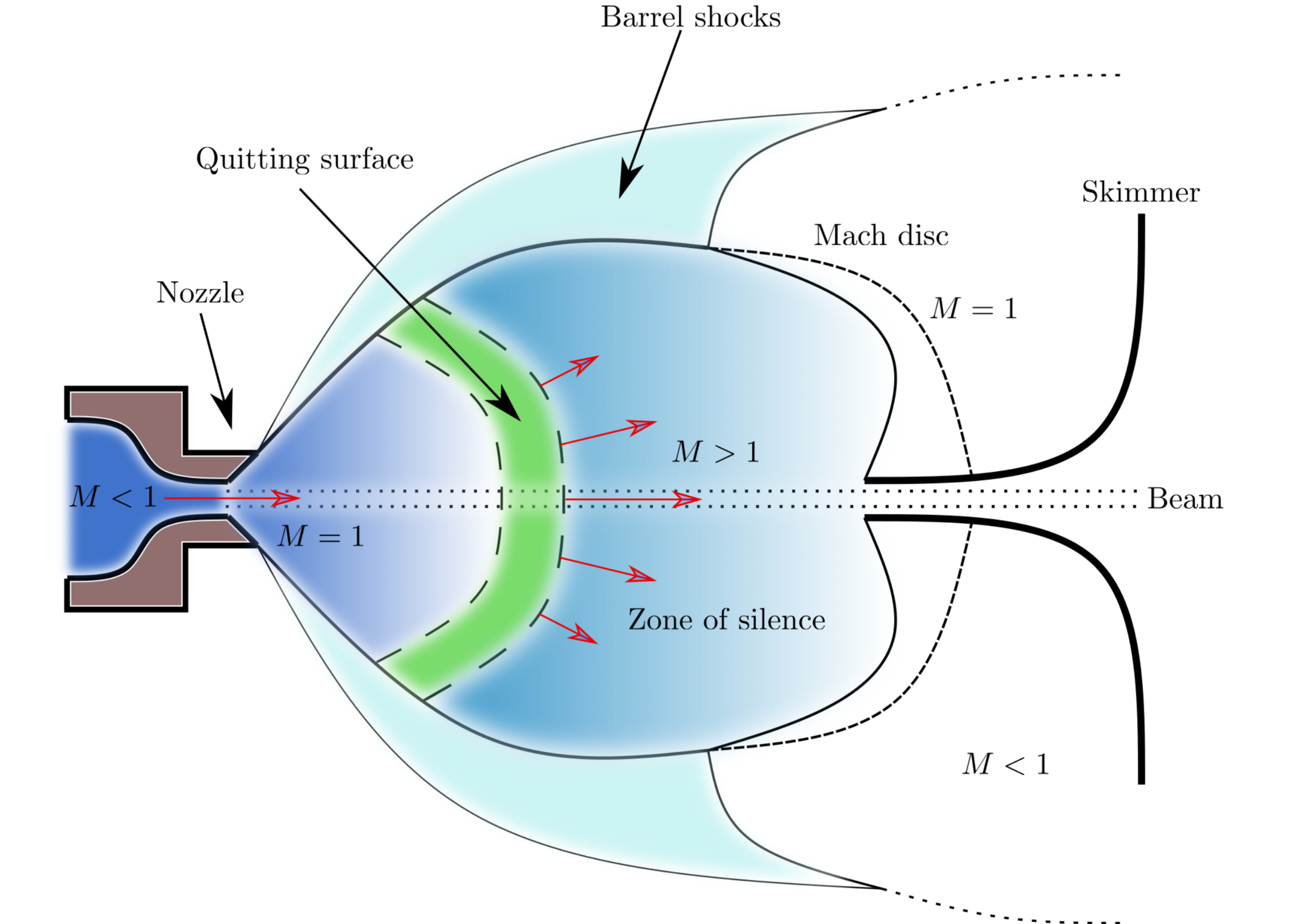}
	\caption{A schematic of a free-jet expansion. The Mach disc and barrel shocks, where the expansion interacts with the background gas in the chamber, are relatively close to the skimmer, representing a typical geometry found in free-jet sources. The skimmer sits inside the zone of silence, within which there is free molecular flow, i.e. the mean free path is large enough for gas-gas interactions to be negligible.}
	\label{free_jet}
\end{figure}

In a typical free-jet source, illustrated by Figure \ref{free_jet}, gas behind a nozzle (typical diameter $d\sim10\ \upmu\mathrm{m}$) is compressed to a high pressure (typically $p_0\sim5-200\ \mathrm{bar}$) \cite{scoles_beam_methods, generalfreejet}. The gas subsequently expands through the aperture at a high velocity, reaching supersonic speeds at the point where the cross sectional area is a minimum, often called the `throat'. As the gas expands into the vacuum beyond, the mean free path increases until it reaches the `quitting surface', an approximate boundary beyond which the mean free path is large enough for gas-gas interactions to be negligible. The region beyond the quitting surface is conventionally referred to as the `zone of silence' \cite{campargue,scoles_beam_methods}. A complicated system of barrel shocks exists at the edges of the zone where the Mach number, $M$, drops below 1 and the expansion interacts with the background gas of the chamber \cite{Khalil2004,Jugroot2004}. The properties of the shock depend on this background pressure, which is primarily determined by the speed of the pump that manages the gas load. In order to develop a beam, a skimmer is positioned in front of the nozzle to extract the center of the expansion, thus defining the beam axis. The skimmer and its mount interact with the forward expansion, leading to backscattering and/or additional shock fronts, which interfere with and attenuate the beam \cite{bird}. The diameter of the skimmer orifice is typically $100\ \upmu\mathrm{m}-1\ \mathrm{mm}$, but microskimmers with smaller apertures exist and constitute an active area of research \cite{microskimmersr,palauintensity}.

The interference is not solely determined by the geometry of the source. If the background pressure, $p_b$, is very low, so that $p_0/p_b = 10^6$ or higher, the source can be described as being in the `Fenn' regime \cite{Yamashita1984,Fenn2000,scoles_beam_methods}. Fenn sources, which are extensively utilized in electrospray ion sources, were enabled by the development of high-throughput diffusion pumps capable of achieving pumping speeds of several thousand $\mathrm{mbar\,L\,s^{-1}}$. In such cases, there is a smooth transition from continuum flow near the nozzle to free molecular flow, which exists throughout most of the chamber. A non-negligible proportion of atoms scatter from the skimmer towards the beam axis, leading to significant interference at cryogenic nozzle temperatures, when all particles move more slowly. The thickness of the Mach disc, which is of the order of the mean-free path, is comparable to the chamber dimension, meaning that complex shock structures are largely absent. In contrast, in the `Campargue' regime, the background pressure is higher ($p_b\gtrapprox0.1\,\mathrm{mbar}$), there exist well defined barrel shocks and the Mach disc sits closer to the nozzle \cite{campargue,oldcamparguemodelling}. As a result, the interference must often be modeled using continuum mechanics and the role of `background' gas is more significant. Accounting for the background pressure is complicated by the fact that the shock fronts are relatively opaque to gas outside the zone of silence \cite{generalfreejet}. A practical solution to this additional interference is to lower the nozzle-skimmer distance, $d_B$, and hence the proportion of the beam scattered by the background gas. However, when $d_B$ is reduced to the extent that it becomes comparable to the diameter of the skimmer orifice, the level of attenuation may rise as the supersonic expansion interacts with and continues within the skimmer itself. This finding can be understood qualitatively from the fact that the number density at the skimmer opening, $n_S$, is approximately proportional to $1/d_B^2$ by the inverse-square law. In particular, if the skimmer is placed within the quitting surface, the gas is underexpanded and complex shock waves can form inside the tip \cite{dsmcbrightening}. So-called `skimmer blocking' may occur for very high flow rates.

There are few published models for source chamber interference and no true analytic ones. The simplest case of a Fenn source has been treated successfully using numerical simulations by Hedgeland et al., although these are not straightforward to run and interpret in practice \cite{anomalous}. While Verheijen et al. explored attenuation with a wedge-shaped skimmer, their theory contains a number of free parameters (e.g. a constant scattering cross section), which they estimate by fitting to experimental data \cite{oldmodelling}. Their method also requires Monte-Carlo simulations for improved accuracy. A related work by Beijerinck et al. applies an analogous method to study the intensity of a Campargue source \cite{oldcamparguemodelling}. Interference within the skimmer itself can be modeled using Direct Simulation Monte Carlo (DSMC) methods \cite{somedsmcsimulations,dsmcbrightening}. For very high number densities, a hybrid approach utilizing continuum fluid mechanics in specific regions may be useful \cite{hybridfluids}. The complexity of these semi-empirical approaches means that it is difficult to quantitatively compare different sources, hampering efforts to improve their design and thus their performance, even when the models are accurate.

In the present work, we focus on sources in which (i) there is free molecular flow and (ii) the nozzle-skimmer distance is large enough for there to be negligible interference within the skimmer itself. Such cases lend themselves to an analytical treatment because the shocks formed are relatively simple. To this end, we show that external skimmer interference can be described by a semi-analytical integral calculation containing no free parameters. Additionally, we determine the relationship between the mean number of interactions, $\mu$, and the measured signal, which justifies the characteristic `turnover' form of many interference curves. We present both cryogenic- and ambient-temperature experimental data to support our conclusions. The results enable us to perform quick and simple calculations of source chamber attenuation, thus addressing the issue of complexity in previous attempts at the problem and obviating the need for simulations. The absence of free or empirical parameters makes our work widely applicable, while the simplicity of the model allows us to derive additional insights. For example, we show that backscattering from the skimmer itself is the principal mode of interference at low nozzle temperatures, whereas in room-temperature sources, the interaction with the background gas dominates. Significantly, we are able to develop a series of design recommendations for skimmers and their mounts, principally for sources operating at cryogenic temperatures. 

\section{Theory}
\label{themaintheory}

In our work, the molecular flow regime is assumed throughout the chamber, i.e. the quitting surface is taken to be very close to the nozzle. We also assume that the free-jet expansion is adiabatic, so we do not distinguish between the stagnation and quitting surface temperatures \cite{scoles_beam_methods,anomalous}.

\subsection{Soft-body scattering cross section}
We first require a scattering cross section, $\sigma$, for the interaction of the gas, which includes both background and backscattered particles, with atoms in the beam. The cross section in long-beam systems has been found empirically to be larger than typical hard-body cross sections in the literature because weak interactions are sufficient to cause deflections of the beam particles away from the axis, preventing transmission through the instrument \cite{anomalous}. A full derivation of the scattering cross section is presented in Appendix \ref{crosssectiondev}, with the key features explained below.

\begin{figure}[h]
	\includegraphics[width=0.495\textwidth]{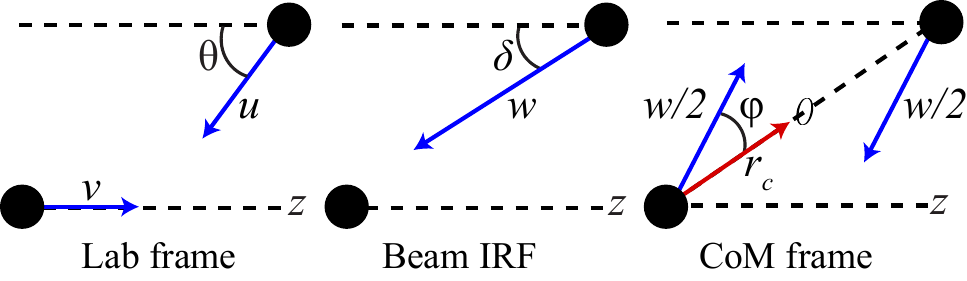}
	\centering
	\caption{Schematics of the three frames needed to derive the scattering cross section between two atoms (black circles). The second and third diagrams correspond to the instantaneous rest frame (IRF) of the beam atom and the center of mass (CoM) frame, respectively.}
	\label{frames}
\end{figure}

Consider a beam particle traveling along the $z$-axis with speed $v$. Let $u$ be the speed of a background gas particle at an angle $\theta$ to the negative $z$-direction. The situation is illustrated in the left-most panel of Figure \ref{frames}. The center of mass (CoM) and instantaneous rest frame (IRF) of the beam atom are also shown. In the latter, the speed of the gas atom is $w=\sqrt{u^2+v^2+2uv\cos\theta}$ and the angle $\delta$ is given by $\tan\delta=u\sin\theta/\left(v+u\cos\theta\right)$.

We approximate the soft-body gas-phase interaction using a van der Waals (vdW) potential, $V\left(r\right)=-C_6/r^6$ for particle separation $r$. Both experimental \cite{scoles_beam_methods} and theoretical \cite{c6valuescomput} values of the coefficient $C_6$ exist; the latter is used in all our calculations. The total deflecting impulse acting on a beam particle can then be evaluated. The maximum possible deflection that still leads to detection is determined by the acceptance angle, $\Omega$, which is, in turn, defined by the limiting aperture along the entire beam length \cite{anomalous}. $\Omega$ is likely to be due to an aperture beyond the skimmer in the majority of instruments; for example, in Sources 1 and 3 discussed later in Section \ref{experimental}, $\Omega$ is due to the pre-detector aperture, whereas in Source 2, its value is dictated by the entrance aperture to the scattering chamber. Combining results, we arrive at

\begin{equation}
	\sigma=\left(\frac{15\pi C_6}{8mvw\Omega}\right)^{1/3}\int_{-\pi/2}^{\pi/2}\left(\cos^2\varphi+\sin^2\varphi\cos^2\delta\right)^{1/6}\mathrm{d}\varphi.
	\label{crosssection}
\end{equation}
 
\subsection{Interaction with backscattered particles}
\begin{figure}[h]
    \centering
    \includegraphics[width=0.9\linewidth]{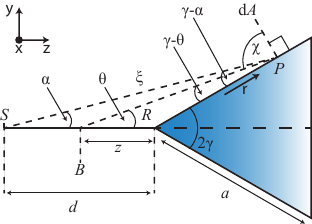}
    \caption{A schematic diagram of a conical skimmer with an opening angle of $2\gamma$ at its tip. A source nozzle is positioned at $S$, $B$ is an arbitrary point on the beam axis between the source and skimmer tip and $P$ is an arbitrary point on the surface of the skimmer.}
    \label{fig:conical_schematic}
\end{figure}

\begin{figure}[h]
	\includegraphics[width=0.9\linewidth]{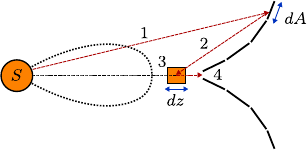}
	\centering
	\caption{A schematic diagram of the steps in our interference calculation. Atoms are emitted from $S$ according to a given distribution (dashed line), leading to a flux at the skimmer ($1$). Some of the particles are scattered towards the beam axis ($2$); they subsequently interfere with it according to a particular cross section ($3$) and slightly perturb the forward trajectories. The beam then passes through the skimmer ($4$) towards the detector. Particles are assumed to travel in straight lines throughout.}
	\label{principle_diagram}
\end{figure}

We begin by defining the geometry and coordinates of the source-skimmer assembly, shown in Figure \ref{fig:conical_schematic}. The corresponding calculation scheme that will be used to evaluate the interference between backscattered or background gas and particles in the beam is depicted in Figure \ref{principle_diagram}.

The scheme splits the problem, which has cylindrical geometry, into conical subsections so that the method can be applied to arbitrary skimmer geometries. For example, a conical skimmer can be approximated by a truncated cone with half-angle $\gamma$ for $r_1<r<r_2$. Following the schematic diagram in Figure \ref{principle_diagram}, we assume that the outgoing flux diverges from a point source at the nozzle (1); in other words, we take the quitting surface to be at the nozzle itself.  Cosine scattering of the expansion from the skimmer delivers a certain fraction back towards the beam axis (which we term backscattered particles) (2) \cite{LambrickDiffuse2022}. Using the scattering cross section derived in Section \ref{crosssection}, the interaction between the beam and the backscattered atoms is calculated (3), leading to small perturbations in the trajectories of atoms along the beam axis that eventually pass through the skimmer (4). We assume that all particles travel in straight lines between scattering events.

Comparing Figures \ref{fig:conical_schematic} and \ref{principle_diagram}, we must first express the distances $\xi=SP$ and $R=BP$ in an appropriate coordinate system. The lengths correspond to calculations of the flux distributions from the nozzle and skimmer, steps (1) and (2) in the calculation scheme, respectively. Due to the inherent cylindrical symmetry of most source chambers, we will use spherical polar coordinates, $(r,\theta,\varphi)$. As $P$ is a point on the skimmer that lies a distance $r$ from its tip, its coordinates can be expressed as $\left(x,y\right)=\left(d+r\cos{\gamma},\  r\sin{\gamma}\right)$. The point $B$ lies on the beam, a distance $z$ from the skimmer tip. $\mathrm{d}A=r\sin{\gamma}\mathrm{d}r\mathrm{d}\varphi$ is the area element of the skimmer at $P$. The distances $\xi$ and $R$ are given by

\begin{gather}
	\xi^2=d^2+r^2+2dr\cos{\gamma},\\ R^2=z^2+r^2+2zr\cos{\gamma}.
	\label{Rxi}
\end{gather}

\noindent From Figure \ref{fig:conical_schematic}, $\cos\alpha=\left(r\cos\gamma+d\right)/\xi$ and $\cos\theta=\left(r\cos\gamma+z\right)/R$. To simplify the following calculations, an assumption must be made regarding the shape of the forwards expansion in Figure \ref{principle_diagram}. Beijerinck and Verster's widely used approximation is:

\begin{equation}
	\frac{\mathrm{d}f}{\mathrm{d}\Omega}=\frac{p+1}{2\pi}\cos^p\alpha,
	\label{fdistribution}
\end{equation}

\noindent where $\mathrm{d}\Omega$ is an element of solid angle and $\text{d}f$ is the probability of scattering into $\mathrm{d}\Omega$ at $\alpha$ \cite{modellingsupersonicdistribution}. A peaking factor of $4$ (i.e. $p=3$, relative to a uniform hemispherical distribution, for which $p=0$) is expected for a monatomic gas, based on a calculation of the flow through a de Laval nozzle \cite{scoles_beam_methods}.

The flux of atoms at $P$ can be calculated from \eqref{fdistribution} and the total flow of particles through the nozzle, $\dot{N}$. Resolving perpendicular to the skimmer's surface to obtain the component of the flux, $F$, incident on $\text{d}A$,

\begin{equation}
	F=\frac{\dot{N}}{\xi^2}\frac{\mathrm{d}f}{\mathrm{d}\Omega}\sin\left(\gamma-\alpha\right)=\frac{\left(p+1\right)Qd\sin\gamma}{2\pi k_BT_0}\frac{\cos^p\alpha}{\xi^3},
	\label{fullflux}
\end{equation}

\noindent where we have used $\sin\left(\gamma-\alpha\right)=d\sin\gamma/\xi$ and $Q=p_0\dot{V}=\dot{N}k_BT_0$ for stagnation pressure $p_0$, nozzle temperature $T_0$ and volumetric flow rate $\dot{V}$. The nozzle throughput, $Q$, is a measurable quantity with units of $\SI{}{\milli\bar\litre\per\second}$.

To evaluate the backscattered flux arriving at $B$ on the beam axis, i.e. step (2) in the calculation scheme, we now consider the scattering of $F$ from the skimmer surface element, $dA$. The diffuse scattering from rough or `unclean' surfaces covered by adsorbates is generally well approximated by a cosine distribution, sometimes biased away from the surface normal \cite{raytracing,image,Eder2023}. As a result, the probability per solid angle of a particle being scattered to an angle $\chi$ can be modeled by

\begin{gather}
	\frac{\mathrm{d}f}{\mathrm{d}\Omega}=\frac{1}{\pi}\cos{\chi}.
	\label{cosinescattering}
\end{gather}

\noindent Although there may be a specular component to the flux scattered from specific materials or very clean surfaces, simulations of the scattering distribution in helium microscopy have shown that a specular portion greater than just $1\%$ is inconsistent with images of adsorbate-covered samples \cite{samcosinescattering}. We therefore do not extend our model to such cases. Now consider scattering towards the beam axis. The rate of particles striking the element $\mathrm{d}A$ is $F\mathrm{d}A$. Hence, the rate of scattering towards $B$ is $F\mathrm{d}A\mathrm{d}f$. Also consider a small area $\mathrm{d}S=R^2\mathrm{d}\Omega$ at $B$, normal to the $\chi$ direction, at which the flux is $\mathrm{d}F_z=F\mathrm{d}A\mathrm{d}f/R^2\mathrm{d}\Omega$. We assume that the attenuation of the backscattered flux by the forwards expansion in the intervening region is negligible. Using $\chi=\pi/2-\left(\gamma-\theta\right)$, $\sin\left(\gamma-\theta\right)=z\sin\gamma/R$ and the expression for $\mathrm{d}A$,

\begin{gather}
	\mathrm{d}F_z=\frac{F\sin^2{\gamma}}{{\pi R}^3}zr\mathrm{d}r\mathrm{d}\varphi.
	\label{fluxz}
\end{gather}

In reality, the particles scattering from the skimmer and source chamber leave with a non-equilibrium distribution of energies that is difficult to predict theoretically. However, a common approximation, first attributed to Maxwell, is to assume that a finite fraction scatter elastically; the remain subset accommodate to the skimmer and chamber temperature, $T_s$, and so attain a Maxwell-Boltzmann speed distribution \cite{thermalaccomodation}. The accommodated fraction, or energy accommodation coefficient, here denoted by $\lambda$, is an empirical measure of the efficiency of energy exchange between a surface and colliding gas phase molecules, averaged over all interactions \cite{lambdadefinition}. In the calculations that follow, we use $\lambda=0.15$, given that the accommodation coefficients for He on adsorbate-covered nickel and stainless steel (of which the skimmer and chamber are composed) are $0.156$ and $\approx0.15$, respectively \cite{thermalaccomodation}. In this way, we follow the approach of Hedgeland et al., which was previously successful in describing interference \cite{anomalous}. However, we note that the results are not substantially different when the value of $\lambda$ is changed; the inclusion of the parameter therefore amounts to a small correction to the scattered flux.

The Maxwell-Boltzmann distribution for the accommodated particles means the probability of leaving with a speed $u\rightarrow u+du$ is

\begin{equation}
	\mathrm{d}f_u=g\mathrm{d}u=4\pi u^2 \left(\frac{m}{2\pi k_BT_S}\right)^{3/2} e^{-mu^2/2k_BT_S}\mathrm{d}u.
	\label{maxwellboltzmann}
\end{equation}

Regarding the elastically scattered particles, the distribution of incoming speeds does not resemble \eqref{maxwellboltzmann}. However, for high Mach numbers, which are achieved in many sources \cite{thesource, bajl, anomalous}, the distribution approaches a delta function. In other words, it is sufficient to assume that $v=v_0$, where

\begin{equation}
	v_0=\sqrt{\frac{2\kappa}{\kappa-1}\frac{k_BT_0}{m}}.
	\label{v0}
\end{equation}

\noindent The above can be derived from an analysis of the flow through a de Laval nozzle \cite{scoles_beam_methods}. $\kappa$ is the adiabatic ratio with a value of $5/3$ for a monatomic gas.

The mean number of interactions can be computed by working in the beam IRF at $B$ and considering particles incident on the element. The number density, $\mathrm{d}n$, associated with $\mathrm{d}F_z$ is invariant, so the flux in the IRF is

\begin{equation}
	\mathrm{d}F_z'=\mathrm{d}nw=\frac{w}{u}\mathrm{d}F_z.
	\label{relativevelocityfactor}
\end{equation}

\noindent The number of interactions experienced by a beam atom in the element of length $\mathrm{d}z= v_0\mathrm{d}t$ is

\begin{equation}
	\mathrm{d}\mu_S=\mathrm{d}F_z'\sigma \mathrm{d}t\mathrm{d}f_u,
	\label{dmusmall}
\end{equation}

\noindent where $\sigma$ is the previously derived scattering cross section. Note that \eqref{dmusmall} is also invariant. The final expression for $\mathrm{d}\mu_S$, expressed as a sum of the elastically scattered and thermally accommodated terms, is

\begin{multline}
	\mathrm{d}\mu_S=\frac{\left(p+1\right)Qd\sin^3\gamma}{2\pi^2k_BT_0v_0}\frac{zrw\sigma\cos^p\alpha}{u\xi^3R^3}\\
	\times\left[\left(1-\lambda\right)\delta\left(u-v_0\right)+\lambda g\right]\mathrm{d}r\mathrm{d}\varphi \mathrm{d}u\mathrm{d}z.
	\label{dmufull}
\end{multline}

Taking \eqref{dmufull} and defining a flow-independent attenuation coefficient due to backscattering, $\eta_S$, by $\mu_S\equiv\eta_S Q$, allows us to write the exact integral form of $\eta_S$ as

\begin{multline}
	\eta_S=\frac{\left(p+1\right)d\sin^3\gamma}{\pi k_BT_0v_0}\int_{0}^\infty\int_{r_-}^{r_+}\int_{d-d_B}^d\frac{zrw\sigma\cos^p\alpha}{u\xi^3R^3}\\
	\times\left[\left(1-\lambda\right)\delta\left(u-v_0\right)+\lambda g\right]\mathrm{d}z\mathrm{d}r\mathrm{d}u.
	\label{etas}
\end{multline}

Following the calculation scheme, we have arrived at an exact expression for the attenuation coefficient due to backscattered particles that is applicable to arbitrary skimmer geometries but cannot be solved analytically. Using Figure \ref{fig:conical_schematic} for reference, progress can be made for a single conical skimmer by assuming that the half-angle $\gamma$ at its tip is small (i.e. $\gamma\ll1$) and taking the radius of the source nozzle aperture at $S$ to be negligible compared to the nozzle-skimmer distance, $d_B$. We can now integrate \eqref{etas} to obtain the following approximate analytic solution

\begin{equation}
    \eta_S\approx\underbrace{\vphantom{\left(\frac{1}{1}\right)}\frac{(p+1)\gamma^3}{16d_Bk_BT_0}}_{\substack{{\text{(a) Free-jet}}\\\text{expansion}}}\overbrace{\left(\frac{15\pi C_6}{mv_0^4\Omega}\right)^{1/3}}^{\substack{\text{(b) Gas-phase}\\\text{scattering}}}\underbrace{\vphantom{\left(\frac{1}{1}\right)^{\frac{1}{3}}}\left[1-\frac{1}{(1+a/d_B)^4}\right]}_{\substack{\text{(c) Skimmer}\\\text{geometry}}}\overbrace{\vphantom{\left(\frac{1}{1}\right)^{\frac{1}{3}}}I_u.}^{\mathclap{\substack{\text{(d) Inelastic}\\\text{effects}}}}
	\label{etasconical_maintext}
\end{equation}

\noindent The equation is composed of four factors; (a) describes the shape of the free-jet expansion from the source aperture, (b) describes the interaction between the backscattered particles and those in the beam while (c) defines the geometry of the source, skimmer and its mount. The final term, $I_u$, arises from gas-gas interactions and is primarily controlled by elastic scattering events (rather than those in which the scattered particles accommodate to the temperature of the chamber). The complete form of $I_u$ is

\begin{equation}
	I_u=\int_0^\infty\frac{\left(u+v_0\right)^{2/3}}{u}\left[\left(1-\lambda\right)\delta\left(u-v_0\right)+\lambda g\right]\mathrm{d}u.
	\label{uspeedintegralexam}
\end{equation}

\noindent For no energy accommodation of the scattered particles, $\lambda=0$ and $I_u=\left(4/v_0\right)^{1/3}$. A crucial finding from \eqref{etasconical_maintext} is that $\eta_S\propto\gamma^3$, so the interference increases very sharply as the angle is made larger. However, the skimmer mount, which is increasingly obscured as $\gamma$ rises, may also need to be taken into account, depending on the extent of the Mach disc.

A comparison between the numerical and approximate analytic solutions for $\eta_S$, \eqref{etas} and \eqref{etasconical_maintext} respectively, is shown graphically in Figure \ref{backscattered_approx}, Appendix \ref{backscat_app}. There is excellent agreement up to relatively large values of $\gamma$. In Appendix \ref{fappendix}, we show how \eqref{etasconical_maintext} can be modified for the case of a generic continuous profile function that describes a skimmer or chamber. In Appendix \ref{backscat_app}, we consider the case of a parabolic skimmer profile as an example, with the results plotted in Figure \ref{parabolic_approximate}.

\subsection{Interaction with background particles}

To find an expression for the interaction of the beam particles with background gas, we consider the lab frame in Figure \ref{frames}, in which the flux of background atoms, $\text{d}F$, is 

\begin{equation}
	\mathrm{d}F=nu\mathrm{d}f_u\mathrm{d}f_\theta.
	\label{initialflux}
\end{equation}

\noindent $\text{d}f_u$ is the fraction of background particles with speeds $u\rightarrow u+\mathrm{d}u$, as in \eqref{maxwellboltzmann}, while $\mathrm{d}f_\theta$ is the probability of a particle moving towards the beam at an angle $\theta\rightarrow\theta+\mathrm{d}\theta$. Assuming that the background gas is completely randomized, $\mathrm{d}f_\theta$ is proportional to the solid angle between $\theta$ and $\theta+\mathrm{d}\theta$. Normalizing, $\text{d}f_{\theta}$ becomes

\begin{equation}
	\mathrm{d}f_\theta=\frac{1}{2}\sin\theta \mathrm{d}\theta,
	\label{dftheta}
\end{equation}
where $n$ is the number density of the background gas. In the sources we consider, $n$ is approximately constant in the source chamber and is given by

\begin{equation}
	n=p_C/k_BT_C,
	\label{idealgaseq}
\end{equation}

\noindent for chamber pressure $p_C=Q/S$, pumping speed $S$ and chamber temperature $T_C$. Equation \ref{idealgaseq} is most accurate in Fenn-type sources, where the weak shock structure is unable to prevent most of the external gas from passing into the zone of silence. In Campargue sources, the stronger shocks mean that $n$ is likely to be smaller. 

Substituting \eqref{dftheta} and \eqref{idealgaseq} into \eqref{initialflux}, we can express the flux of background atoms in the lab frame as

\begin{equation}
    \text{d}F = \frac{uQ\sin{\theta}}{2Sk_BT_C}\text{d}f_u \text{d}\theta.
    \label{dF_pre_frame_trans}
\end{equation}

We now transform \eqref{dF_pre_frame_trans} from the lab frame into the beam IRF using \eqref{relativevelocityfactor}, to arrive at

\begin{equation}
    \text{d}F'= \frac{wQ\sin{\theta}}{2Sk_BT_C}\text{d}f_u \text{d}\theta.
    \label{dF_post_frame_trans}
\end{equation}

\noindent From \eqref{dmusmall}, the mean number of scattering events between beam and background particles is $\text{d}\mu_B=\text{d}F'\sigma \text{d}t$, so that \eqref{dF_post_frame_trans} becomes

\begin{equation}
    \text{d}\mu_B=\frac{wQg\sigma\sin{\theta}}{2Sk_BT_Cv_0}\text{d}\theta \text{d}u \text{d}z,
    \label{dmu}
\end{equation}

\noindent which is analogous to \eqref{dmufull} for the case of interaction between an atom in the beam and backscattered particles. By definition, the mean number of scattering events is related to the attenuation coefficient of the beam by $\mu_B\equiv\eta_BQ$. Applying this relation to \eqref{dmu} and integrating gives

\begin{equation}
	\eta_B=\frac{d_B}{2Sk_BT_Cv_0}\int_{0}^\infty\int_{0}^\pi w\sigma\sin\theta g \mathrm{d}\theta \mathrm{d}u.
	\label{etab}
\end{equation}

The integral for $\eta_B$ cannot be evaluated analytically. However, for nozzle temperatures that are much lower than the chamber temperature, $T_0\ll T_C$, the interfering particles are travelling much slower than those in the beam, so $u\gg v_0$. As a result, the relative speed of the background particles satisfies $w\approx u$, allowing us to arrive at

\begin{equation}
	\eta_B\approx\overbrace{\vphantom{\left(\frac{1}{1}\right)^{\frac{1}{3}}}\frac{5\Gamma\left(5/6\right)d_B}{12S}}^{\substack{\text{(a) Effective}\\\text{path length}}}\underbrace{\vphantom{\left(\frac{1}{1}\right)^{\frac{1}{3}}}\left[\frac{30C_6}{\sqrt{\pi}\Omega\left(mk_BT_C\right)^2v_0^4}\right]^{1/3}}_{\substack{\text{(b) Scattering } \text{cross section}}}\overbrace{\vphantom{\left(\frac{1}{1}\right)^{\frac{1}{3}}}I_\theta,}^{\mathclap{\substack{\text{(c) Inelastic}\\\text{effects}}}}
	\label{etabapprox}
\end{equation}

\noindent where $\Gamma$ is the gamma function. The approximate solution can be broadly described as consisting of three components. Part (a) describes the effective path length of a beam particle through the interfering background gas, (b) accounts for the soft-body vdW potential for gas-phase scattering that determines the scattering cross section, while (c) encapsulates the inelastic scattering effects within a function $I_{\theta}$, a numerical constant calculated as

\begin{equation}
	I_\theta=\int_0^\pi\int_{-\pi/2}^{\pi/2}\left(\cos^2\varphi+\sin^2\varphi\cos^2\theta\right)^{1/6}\sin\theta \mathrm{d}\varphi \mathrm{d}\theta\approx5.72.
	\label{horriblenumericalintegral}
\end{equation}

For completeness, we plot the exact and approximate expressions for $\eta_B$ in Figure \ref{background_approximate}, Appendix \ref{background_app} using $T_C=293\ \mathrm{K}$ and $S=2500\ \mathrm{L\,s^{-1}}$, showing excellent agreement between the solutions. However, a different pumping speed simply leads to a re-scaling of all values of the coefficient, because $\eta_B\propto1/S$.

\subsection{Relationship between attenuation coefficient and measurable signal}

The mean number of interactions that deflect a particle from the beam is simply a sum of the terms due to direct backscattering and background gas, i.e $\mu=\mu_S+\mu_B$. Each is proportional to $Q$, so the total attenuation coefficient is $\eta=\eta_S+\eta_B$. 

To find a relationship between $\mu$ and the signal, $W$, Poisson statistics can be applied, given that disruptive interactions are independent events in the limit of small-angle scattering. Thus, the probability that a particle suffers $n$ collisions is $P\left(n\right)=\mu^ne^{-\mu}/n!$. The probability of zero events, which must be the same as the detected fraction of the beam, is $P\left(0\right)=e^{-\mu}$. The signal is proportional to the nozzle throughput in the absence of interference, so

\begin{equation}
	W\propto Qe^{-\eta Q},
	\label{W}
\end{equation}

\noindent which implies that the signal is maximal for a throughput of $Q_M=1/\eta$. The top panels of Figure \ref{fig:big_figure} show that \eqref{W} models experimental data from literature very well. The fact that the signal can be described by $Qe^{-\eta Q}$ is significant, irrespective of our theory to calculate $\eta$, because it means that source chamber interference can be quantified by a single parameter.

However, many signal-throughput curves do not resemble the form given by \eqref{W} \cite{palauintensity}. A plateau in signal is commonly observed, instead of a turnover. The majority of such deviations can be attributed to the use of high-pressure (i.e. non-Fenn type) sources, in which problematic shock waves at the skimmer entrance develop. Studies have shown that cooling the skimmer to suppress such shock waves may completely eliminate the interference \cite{dsmcbrightening}. Once this so-called skimmer `blocking' is accounted for, \eqref{W} is still applicable.

\subsection{Evaluating the scattering coefficient of a chamber}
In the calculations that follow, we consider an axisymmetric vacuum chamber described in profile by coordinates $\bm{x}$ and $\bm{y}$, such that an individual element is represented by $\left(x_i,\,y_i\right)\rightarrow\left(x_{i+1},\,y_{i+1}\right)$. Each element is described by an angle $\gamma_i$, an effective nozzle-tip distance $d_i$ and limits of integration $r\in[r_i^-,r_i^+]$, where

\begin{equation}
	\gamma_i=\tan^{-1}{\left(\frac{y_{i+1}-y_i}{x_{i+1}-x_i}\right)},
	\label{gammai}
\end{equation}

\begin{equation}
	d_i=\frac{y_{i+1}x_i-y_ix_{i+1}}{y_{i+1}-y_i},
	\label{di}
\end{equation}

\begin{equation}
	r_i^-=y_i\sqrt{1+\left(\frac{x_{i+1}-x_i}{y_{i+1}-y_i}\right)^2},
	\label{yiminus}
\end{equation}

\begin{equation}
	r_i^+=y_{i+1}\sqrt{1+\left(\frac{x_{i+1}-x_i}{y_{i+1}-y_i}\right)^2}.
	\label{yiplus}
\end{equation}

Geometrical shadowing may also occur in two ways. First, shadows cast by particular elements prevent others from being illuminated by the forward expansion. The occluded elements therefore cannot contribute to the attenuation. Second, particles scattered towards the beam axis may be blocked by intervening elements. As a result, the length of the region in which interference may occur is reduced at certain scattering points. The effect can be accounted for by making $d_B$, which represents the effective beam length, a function of  $r$.

The Mach disc, beyond which the flow ceases to be supersonic, must also be considered. Approaching the disc, $M\rightarrow1$ and the number density rises rapidly, meaning that molecules reaching the boundary interact with it and have their velocities randomized. As a result, the disc is partially opaque to the supersonic expansion and so any elements of the source chamber lying behind it contribute less to $\eta_S$ \cite{scoles_beam_methods}. The position of the disc is given by

\begin{equation}
	x_M=0.67D\sqrt{\frac{p_0}{p_c}}.
	\label{Machdisclocation}
\end{equation}

\noindent Here, $D$ is the nozzle diameter. The dimensionless coefficient of \eqref{Machdisclocation} is empirical, although the result has a sound theoretical basis \cite{Machdisctheory}. There is experimental evidence to suggest that excluding parts of the source beyond the Mach disc is valid \cite{bajl}. We therefore include the disc in our calculations, and neglect the contribution of any elements beyond it to $\eta_S$.

Once a routine to account for shadowing and the Mach disc is implemented, one can calculate $\eta_S$ for each element, before obtaining the total. An alternative approach, in which the skimmer is described by a continuous profile function $f\left(x\right)$, is given in Appendix \ref{fappendix}. 

\subsection{Scaling relationships}
\label{scaling}

Both $\eta_S$ and $\eta_B$ rise as the nozzle temperature (and hence beam momentum) falls. For $\lambda=0$, $\eta_S$ obeys

\begin{equation}
	\eta_S\propto1/T_0^{11/6}.
	\label{etasprop}
\end{equation}

\noindent The proportionality relationship for $\eta_B$ is more complicated as $T_0$ appears in the integrand of \eqref{etab}. However, for chamber temperature $T_C\gg T_0$ equation \eqref{etabapprox} gives,

\begin{equation}
	\eta_B\propto1/T_0^{2/3}.
	\label{etabprop}
\end{equation}

\noindent $\eta_B$ therefore depends more weakly on nozzle temperature than $\eta_S$. The difference arises from the fact that scattering by the background gas molecules is determined by $T_C$, whereas $T_0$ dictates the speed of the majority of the particles that contribute to $\eta_S$. Therefore, because $\eta_B$ decreases less than $\eta_S$ as the nozzle temperature $T_0$ rises, \eqref{etabprop} and \eqref{etasprop} together imply that there is a transition from `skimmer interference' to background-dominated attenuation. We discuss experimental evidence for the phenomenon in Section \ref{experimental}. The existence of a transition implies that there is limited benefit to carefully optimizing the design of skimmer geometries used in ambient-temperature sources.

The dependence of each attenuation coefficient on the nozzle-skimmer distance, $d_B$, is also worth discussing. The background coefficient, $\eta_B$, obeys

\begin{equation}
	\eta_B\propto d_B,
	\label{etabpropd}
\end{equation} 

\noindent given that it represents Beer's law interference. However, the backscattering attenuation coefficient, $\eta_S$, drops off rapidly as $d_B$ increases. For large $d_B$,

\begin{equation}
	\eta_S\propto1/d_B^2,
	\label{etaspropd}
\end{equation}

\noindent which follows from the inverse-square law for the flux of the supersonic expansion arriving at the skimmer. Furthermore, because the two coefficients vary oppositely with $d_B$, the theory predicts a distance at which the interference is minimal ($\partial\eta/\partial d_B=0$). Such a maximum in signal is observed experimentally \cite{scoles_beam_methods,bajl}. Even so, a calculation of the optimum position using the theory here is unlikely to be accurate for all sources. Interference within the skimmer itself, which is best modeled using DSMC methods due to the complex shock structures involved, may contribute for high throughputs and small nozzle-skimmer distances \cite{dsmcbrightening}.

\section{Experimental data}
\label{experimental}

Here we present three sets of interference curves, corresponding to source chambers operating in different regimes, to support our theory. Each is fitted to $W\propto Qe^{-\eta Q}$, which is applicable in all cases. The first two datasets (Subsections \ref{scatteringdom} and \ref{backdom}) are novel, whereas the last (Subsection \ref{transitionevidence}) is from the literature \cite{anomalous}. The attenuation coefficients, $\eta$, are modeled using Equations \eqref{etas} and \eqref{etab}. The throughput, $Q$, is calculated from the stagnation pressure, $p_0$, via the empirical relationship

\begin{equation}
	Q=C\sqrt{\frac{T_C}{T_0}}p_0D^2
	\label{Qfromp0}
\end{equation}

\noindent which has been verified for the relevant sources \cite{andy,bajl}. For \textsuperscript{4}He, $C=450\ \mathrm{ms^{-1}}$ \cite{scoles_beam_methods}. A theoretical expression for $Q$, derived by considering the fluid mechanics of a converging-diverging de Laval nozzle, reveals that $C\propto1/\sqrt{m}$ \cite{scoles_beam_methods}. Hence, $C$ for \textsuperscript{3}He is $\sqrt{4/3}$ times the \textsuperscript{4}He value.

\begin{figure*}[t]
    \centering
    \includegraphics[width=0.99\textwidth]{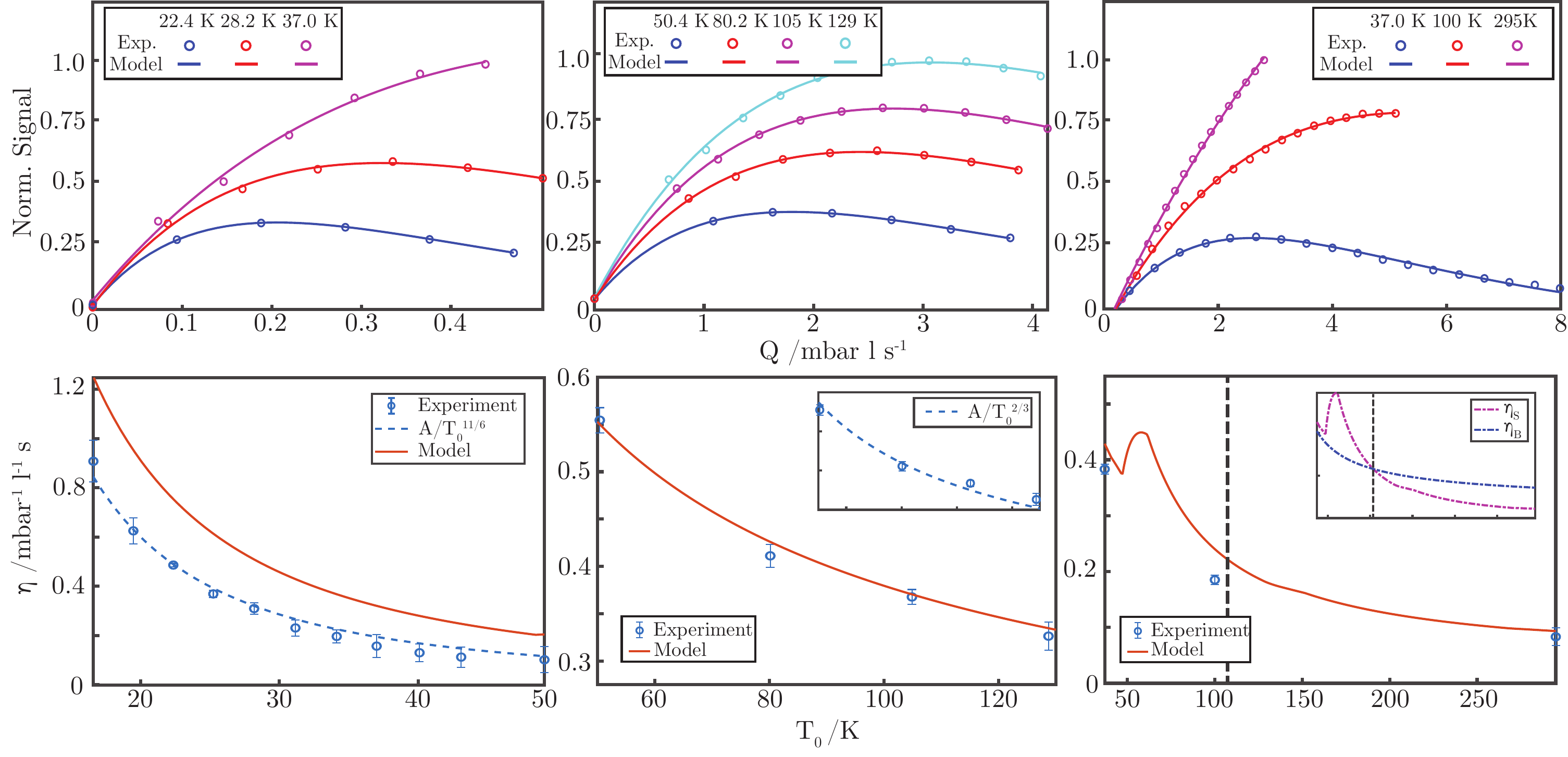}    
    \caption{We demonstrate the accuracy of our model by applying it to Sources 1-3 (left to right), which operate in different pressure regimes and with differing source geometries (shown in Figure \ref{skimmer_geometries_figure}). The cases are dominated by backscattering (left), background (center), and neither (right). In the latter case, a transition between the two interference mechanisms is marked by a vertical dashed line. Top panels - all signal-flow curves show excellent agreement with the model function, $W\propto Qe^{-\eta Q}$. Signal values have been normalized. Bottom panels - the total attenuation coefficient, $\eta=\eta_S+\eta_B$, agrees well with the model in all cases, in both absolute terms and with respect to the scaling relationships predicted by the theory. The experimental data in the rightmost panels are from Hedgeland et al.}

    \label{fig:big_figure}
\end{figure*}

\subsection{A cryogenic source - backscattering-dominated attenuation}
\label{scatteringdom}

We first consider a conical $^3$He source operating with cryogenic nozzle temperatures ($22-37\,\mathrm{K}$), henceforth referred to as Source 1 \cite{thesource}, in which the nozzle-skimmer distance is $6.2\ \mathrm{mm}$ and the pumping speed is $S\approx3000\ \mathrm{L\,s^{-1}}$. A schematic of the source-skimmer geometry is depicted in Figure \ref{skimmer_geometries_figure}. The source contains a $D=10\ \upmu\mathrm{m}$ nozzle and a limiting aperture along the beam axis with an angular radius of $\Omega=1.366\times10^{-3}\ \mathrm{rad}$. The signal reflected from a sample at specular orientation in the scattering chamber was recorded using a custom He atom detector \cite{chisnall}. A selection of the signal-flow curves and both experimental and theoretical values of $\eta$ are shown in the leftmost panels of Figure \ref{fig:big_figure}. No free parameters were used in the modeling of $\eta$ with Equation \eqref{etas}. For simplicity, the background attenuation has been neglected, given that the low beam temperatures and small nozzle-skimmer distance ensure that $\eta_B$ is substantially less than $\eta_S$. Overall, there is reasonable agreement, and the approximate relationship $\eta_S\propto1/T_0^{11/6}$ (dashed line) models the experimental data well. The results constitute strong evidence that for low nozzle temperatures, backscattering is the dominant interference mechanism. 

\subsection{A high-pressure source - background-dominated attenuation}
\label{backdom}

We also consider a $^4$He source, 2 \cite{bajl}. Although similar stagnation temperatures are used here and in the other sources, they operate in different background pressure regimes; the pumping speed on Source 2 (estimated at $1200\ \mathrm{L\,s^{-1}}$) is considerably lower and the nozzle-skimmer distance ($d_B=12\ \mathrm{mm}$) is also larger. As a result, the attenuation of the beam is dominated by the diffuse background gas. The source contains a $D=10\ \upmu\mathrm{m}$ nozzle and the limiting aperture along the beam axis, with an angular radius of $\Omega=1.93\times10^{-3}\ \mathrm{rad}$, lies just before the scattering chamber. The geometry of the source otherwise has an insignificant impact on the attenuation. A selection of signal-flow curves, measured using a Hiden mass spectrometer, is displayed in the top-center panel of Figure \ref{fig:big_figure}. The extracted values of $\eta$ are plotted in the bottom center panel of Figure \ref{fig:big_figure}, alongside the theoretical result of \eqref{etab}. For simplicity, the attenuation due to the skimmer and mount has been neglected, as $\eta_S\ll\eta_B$. There is excellent agreement and the approximate relationship $\eta_B\propto1/T_0^{2/3}$ (dashed line) models the experimental data well.

\subsection{Evidence for a transition in the interference mechanism}
\label{transitionevidence}

Finally, we consider Source 3, for which the theoretical and experimental data from Hedgeland et al. have been used \cite{anomalous}. The data constitute a severe test of the theory, given that the nozzle temperature varies over an order of magnitude, from $27-295\,\mathrm{K}$. Source 3 uses \textsuperscript{4}He and its source-skimmer geometry is shown in Figure \ref{skimmer_geometries_figure}, Panel (b). The nozzle-skimmer distance is relatively large ($15\ \mathrm{mm}$) as a diffusion pump with a speed of $S=4500\ \mathrm{L\,s^{-1}}$ is installed on the chamber, making background attenuation less problematic. The diameter of the nozzle is $18.5\ \upmu\mathrm{m}$ and the limiting aperture was found to produce a beam with an angular radius of $\Omega=2.74\times10^{-3}\ \mathrm{rad}$. In the rightmost panels of Figure \ref{fig:big_figure}, the experimental and theoretical values of $\eta$ are plotted as a function of the stagnation temperature, $T_0$. Both contributions to the calculated coefficient, corresponding to \eqref{etas} and \eqref{etab}, are graphed separately in the inset. There is good agreement, and a transition from backscattering- to background-dominated interference is clearly visible, marked by a dashed black line in the bottom-right panel of Figure \ref{fig:big_figure}, supporting the theory in Subsection \ref{scaling}.

A notable feature of the bottom-right panel in Figure \ref{fig:big_figure} is the dip and rise in the coefficient as the nozzle temperature increases near $37\ \mathrm{K}$, which can be attributed to $\eta_S$ rising as the the Mach disc moves backwards (because $x_M\propto T_0^{1/4}$). As a result, more elements of the source chamber become exposed to the expanding gas and thus contribute to the attenuation by backscattering. In reality, the feature is not observed because the Mach disc is poorly modeled by a sharp boundary where it interacts with the skimmer and chamber \cite{oldcamparguemodelling}. A more accurate, diffuse model for the disc would lead to a smoothing of $\eta$ at low nozzle temperatures.

\section{Source chamber design optimization}
One may use the attenuation coefficient, $\eta$, to minimize interference in, and thus optimize, the design of skimmers in Fenn-type sources. Here, we discuss some general principles based on our theory, alongside a discussion of the optimal skimmer design. We also qualitatively discuss other types of interference that cannot necessarily be ignored. It should be noted that a small drop in $\eta$ corresponds to a substantial increase in signal for the same throughput, $Q$, as the attenuation depends exponentially on the coefficient. Quantitatively, a reduction in its value of $\Delta\eta$ leads to the signal being $e^{\Delta\eta Q}$ times as large.

\subsection{Principle 1: the skimmer design in room-temperature sources is not critical}
\label{designdoesntmatter}

In typical Fenn-type sources operating with room-temperature nozzles, the design of the skimmer and chamber should have a negligible impact on the beam intensity, and any attenuation should principally be due to the background gas. Conversely, in cryogenic sources, there is a diminishing benefit to installing more effective pumps to reduce the chamber pressure because backscattering is more important. Consequently, there should be little difference between a simple conical skimmer, which could be cheaply and rapidly 3D-printed \cite{Radic_plastics_2023}, and a flared, polished design which necessitates advanced tooling.

For room-temperature beams, our theory predicts that (i) reducing the nozzle-skimmer distance $d_B$ and (ii) increasing the pumping speed $S$ have the greatest positive impact on the transmitted intensity. Although increasing $S$ is generally beneficial, reducing the nozzle-skimmer distance too much may lead to internal interference and/or the formation of shock waves. However, cooling the skimmer to $\sim30\ \mathrm{K}$ has been shown to suppress such shock waves, with the precise chamber temperature dependent on the gas utilized \cite{dsmcbrightening}. Further discussion of shocks and their suppression is beyond the scope of the present work.

Henceforth, the discussion will only apply to cryogenic sources, in which attenuation due to backscattering from the skimmer and mount is not negligible.

\subsection{Principle 2: avoid perpendicular surfaces}
\label{noflatmounts}

The benefits of a curved skimmer, with a shallow angle to the beam axis along most of its length, are well understood on the basis of cosine scattering \cite{LambrickDiffuse2022}. However, many skimmer mounts are designed with plane surfaces perpendicular to the beam and/or expansion. Given that such components are not necessarily shielded by the Mach disc, it is important to consider more than just the skimmer when minimizing interference.

\begin{figure}[h]
	\includegraphics[width=0.5\textwidth]{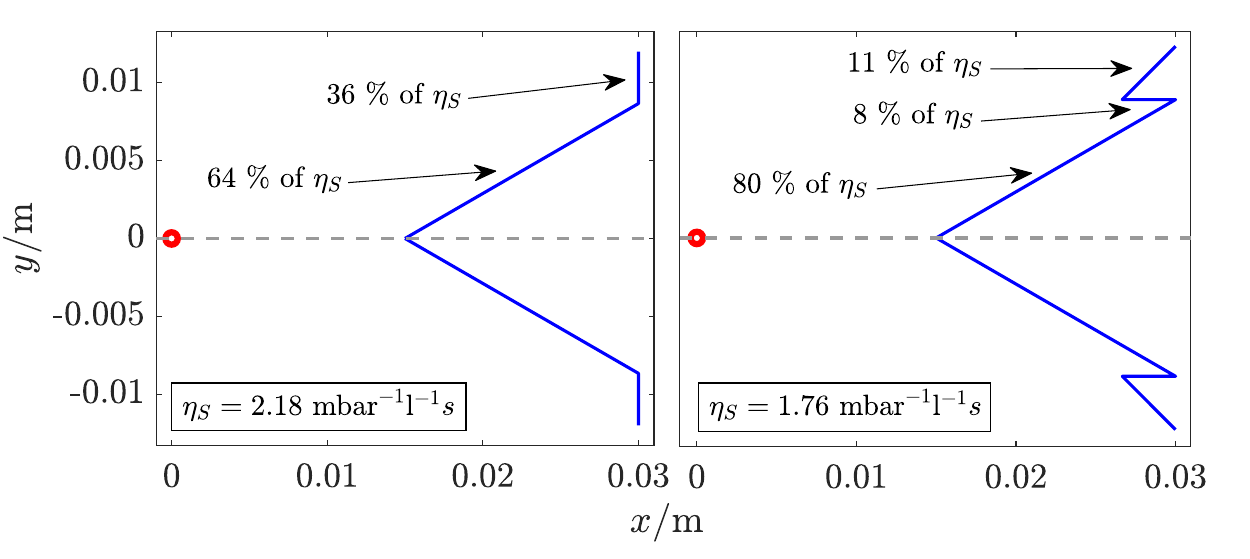}
	\centering
	\caption{A conical skimmer with (i) a conventional flat mount (left) and (ii) a mount angled at $45\degree$ to the beam axis (right). The angled mount reduces the backscattering attenuation coefficient by $\sim\SI{20}{\percent}$ for the same skimmer.}
	\label{original_chamber_mod}
\end{figure}

To illustrate the principle, we consider a \textsuperscript{4}He source, depicted in Figure \ref{original_chamber_mod}, consisting of a conical skimmer and mount. We assume that the pumping speed is large, such that the Mach disc lies behind the arrangement. In the calculations of $\eta_S$, the results of which are stated in the image, we take $\Omega=1.366\ \mathrm{rad}^{-1}$ and $T_0=37\ \mathrm{K}$. Scattering from the skimmer accounts for $64\ \%$ of the attenuation coefficient, meaning that the mount contributes the remaining $36\ \%$, despite it being much smaller and positioned further from the beam. The proportions are almost temperature independent. We then consider the same skimmer with a new mount, such that its face is at $45\degree$ to the beam axis, which is also shown in Figure \ref{original_chamber_mod}. The mount is projected outwards to avoid interfering with the brim of the skimmer. The total contribution of the mount to the attenuation coefficient drops to $20\ \%$, and $\eta_S$ is significantly less.

\subsection{Principle 3: skimmer apertures are problematic}

\begin{figure}[h]
	\includegraphics[width=0.35\textwidth]{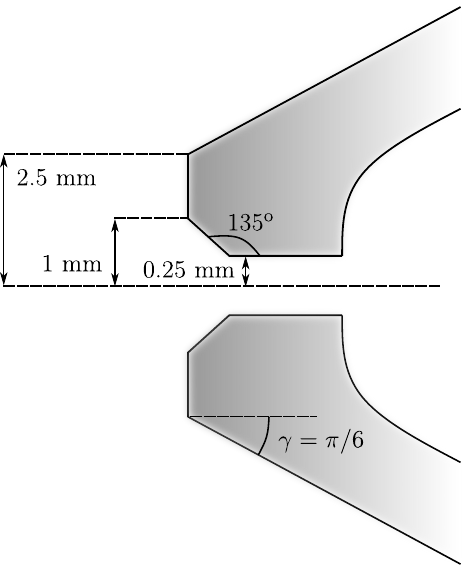}
	\centering
	\caption{A schematic of the tip of an indented skimmer, with the relevant dimensions indicated. The rest of the skimmer is conical, with a length of $15\ \mathrm{mm}$ and no mount. We take the nozzle-skimmer distance to be $d_B=15\ \mathrm{mm}$ in the calculations.}
	\label{fig:indented_skimmer}
\end{figure}

Further to the discussion Subsection \ref{noflatmounts}, our theory suggests that inserting an aperture or glass microskimmer into the end of a standard skimmer is likely to result in significantly more attenuation. An element near the beam axis naturally leads to more interference, by the inverse-square law for flux involved, and perpendicular faces - such as those of an aperture - are highly problematic. A number of designs, including microskimmers, have been tested in the context of intensity and brightness measurements by Palau et al. \cite{adria}. They also consider a skimmer with an indented tip specifically designed to increase attenuation so that it can be studied, henceforth referred to as the `indented skimmer', shown in figure \ref{fig:indented_skimmer}. Although their measurements were recorded for relatively high stagnation temperatures $(\sim300\,\mathrm{K}$ and $\sim125\,\mathrm{K})$ they still conclude that skimmer interference is significant due to the tip geometry.

Here we illustrate the point by comparing a typical conical skimmer to an indented tip design similar to that of Palau et al. \cite{adria}. In our calculations, the two skimmers are identical aside from the tip itself. The length of each is taken to be $l=15\ \mathrm{mm}$, we assume a nozzle-skimmer distance of $d_B=15\ \mathrm{mm}$ and neglect the skimmer mount for simplicity. We consider a \textsuperscript{4}He source at $T_0=37\ \mathrm{K}$ and take $\Omega=1.366\times10^{-3}\ \mathrm{rad}$. For such low nozzle temperatures, the results are striking; $\eta_S=12.5\ \mathrm{mbar\,L\,s^{-1}}$ for the indented skimmer while $\eta_S=0.391\ \mathrm{mbar\,L\,s^{-1}}$ for the equivalent conical skimmer. In other words, the perpendicular face and indent of the skimmer considered by Palau et al. result in overwhelming interference, and contribute $50\ \%$ and $46\ \%$ to the total value of $\eta_S$, respectively. The calculations support the widely accepted principle that skimmer orifices must be sharp and thin. Although the work by Palau et al. is qualitative evidence in support of our results, deviations from our theory are expected if the structure of the skimmer tip leads to blocking and the formation of shock waves.

Further calculations reveal that $\eta_S$ and $\eta_B$ are comparable for room-temperature beams. Therefore, although indented skimmers are impractical with cryogenic nozzles due to the resulting interference, they are still usable in ambient-temperature sources. By extension, glass microskimmers - which also have large values of $\eta_S$, although not to the same extent as indented skimmers - are expected to be acceptable in room-temperature sources but problematic otherwise.

\subsection{Principle 4: a narrow skimmer is not necessarily superior}

There are two reason why a narrow skimmer is not necessarily superior to one with a broader base. The first has been alluded to in Subsection \ref{designdoesntmatter}; internal interference may be problematic in narrow skimmers, especially when the gas is underexpanded and the number density is high at the orifice \cite{dsmcbrightening,somedsmcsimulations}. A minimum opening angle at the skimmer aperture may therefore be needed to prevent `blocking'. The second relates to skimmer mount; as the skimmer is made narrower, more of the chamber behind it is exposed to the forwards expansion. As discussed in Subsection \ref{noflatmounts}, the resulting contribution to the attenuation may be significant and overcome the benefit of sharpening the skimmer.

\begin{figure}[H]
	\includegraphics[width=0.5\textwidth]{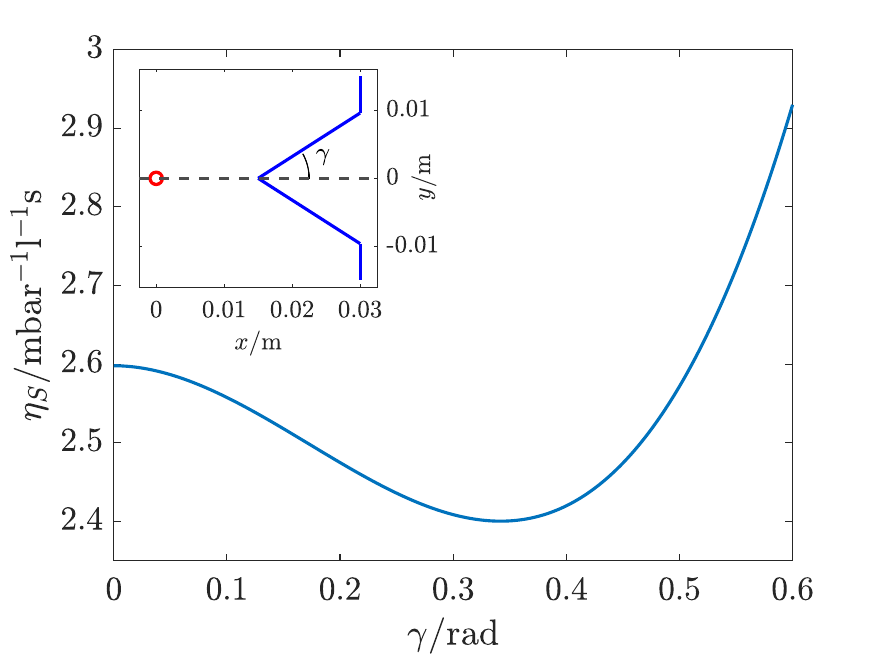}
	\centering
	\caption{A plot of the backscattering attenuation coefficient, $\eta_S$, as a function of $\gamma$ for a skimmer (length $15\ \mathrm{mm}$) and mount (radius $25\ \mathrm{mm}$). As the skimmer broadens, the total interference due to backscattering drops because the skimmer shields an increasing fraction of the mounting plate. The attenuation subsequently reaches a minimum value and rises.}
	\label{mount_minimum}
\end{figure}

We consider a simple system to illustrate the point, depicted in the inset of Figure \ref{mount_minimum}, consisting of a conical skimmer and mount. However, the conclusions generalize to curved skimmers and more complicated geometries. We vary the angle, $\gamma$, at the skimmer opening and plot the results in Figure \ref{mount_minimum}. We otherwise take the parameters from Source 3 in in Subsection \ref{transitionevidence}, i.e. we consider \textsuperscript{4}He at $T_0=37\ \mathrm{K}$ with a representative value of $\Omega=1.366\times10^{-3}\ \mathrm{mbar}$. We also ignore background attenuation and the Mach disc. There is a clear minimum in the value of the attenuation coefficient at $\gamma = \SI{0.35}{\radian}\,(\approx \SI{20}{\degree})$. By comparison, the most popular and widely available commercial skimmers from Beam Dynamics Inc. have opening half-angles of $\gamma=\SI{12.5}{\degree}$ for their stock units (namely Models 1 and 2 \cite{beamDynamicsHomepage, beamDynamicsSkimmerSpecs}). These values of $\gamma$ would result in the backscattering attenuation coefficient being approximately $5\%$ than our suggested optimal value, according to Figure \ref{mount_minimum}, if all other dimensions are the same as modeled here.

\subsection{Optimization of skimmer design}

Here we calculate the optimum value for the radius of the circle describing the skimmer arc, after fixing its length and width. We stress that current skimmers are perfectly usable and the calculations here serve as a demonstration of the theory. In this example, we take the nozzle-skimmer distance, skimmer length and skimmer width to be $d_B=15\ \mathrm{mm}$, $l=15\ \mathrm{mm}$ and $w=10\ \mathrm{mm}$, respectively. We then vary the radius of curvature, $R_C$; a sharper skimmer corresponds to a smaller value of $R_C$ whereas a conical skimmer is recovered as $R_C\rightarrow\infty$. Naturally, when designing a real source, all four parameters can be varied and thus must be considered alongside other constraints. 

\begin{figure}[h]
	\includegraphics[width=0.5\textwidth]{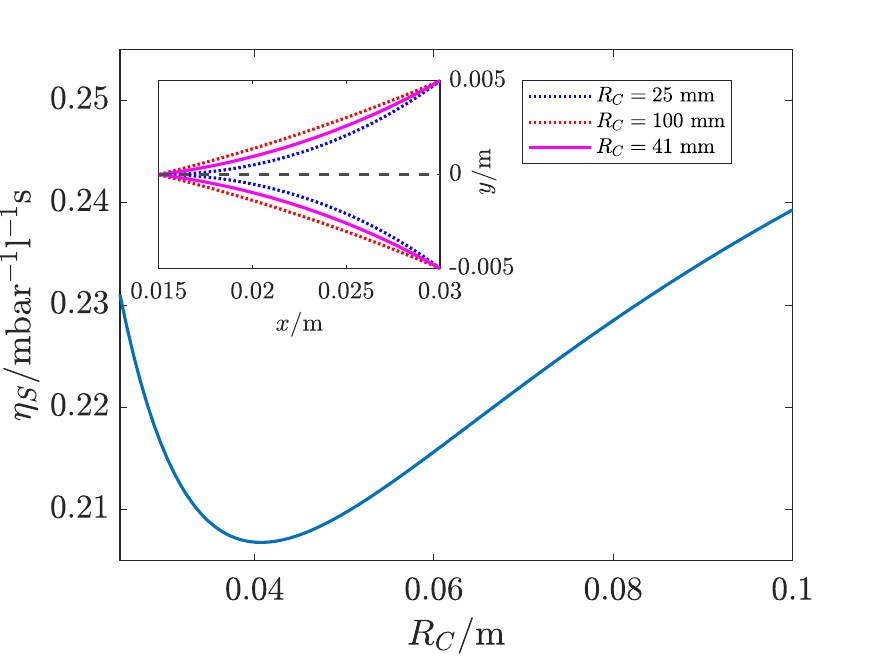}
	\centering
	\caption{A plot of the backscattering attenuation coefficient, $\eta_S$, for a curved skimmer with a profile described by an arc section with a radius $R_C$. The arc is constrained to pass through the beam axis. The width of the skimmer base is fixed at $w=10\ \mathrm{mm}$, its length is $l=15\ \mathrm{mm}$ and the nozzle-skimmer distance is $d_B=15\ \mathrm{mm}$. A broad minimum is observed, showing that the theory can be applied to optimize skimmer designs.}
	\label{optimal_design}
\end{figure}

The results are plotted in Figure \ref{optimal_design}. We consider a \textsuperscript{4}He source at $T_0=37\ \mathrm{K}$, take $\Omega=1.366\times10^{-3}$ and neglect background attenuation. Two cases and the optimal design are shown in the inset. Crucially, the results show that a sharper skimmer is not necessarily superior to one with a flatter profile. The optimal radius of curvature, corresponding to the minimum value of the attenuation coefficient, is $\SI{41}{\milli\metre}$.

\section{Conclusions}

We have described a method for calculating the impact of source chamber interference on the intensity of a free-jet beam. Our results are applicable to Fenn sources, in which the chamber pressure does not completely suppress free molecular flow. Our approach is more straightforward and complete than previous attempts at the problem, as we have shown that the beam attenuation can be described by a single parameter, $\eta$, given by an integral expression containing no adjustable parameters. The signal, $W$, is proportional to $Qe^{-\eta Q}$ for nozzle throughput, $Q$. $\eta$ is the sum of two components, due to backscattering and the interaction with the background gas, respectively. Equations \eqref{etas} and \eqref{etab}, for $\eta_S$ and $\eta_B$, summarize the content of the work and are simple to evaluate numerically. We have also presented three sets of experimental data in Section \ref{experimental} to support the results, with good agreement in all cases, particularly with respect to the scaling relationships ($\eta_S\propto1/T_0^{11/6}$ and $\eta_B\propto1/T_0^{2/3}$) predicted by our theory.

We have shown that `skimmer interference' is only significant for typical source-skimmer geometries when the nozzle temperature is low, whereas the interaction with the background gas is the dominant contribution to the attenuation of room-temperature beams. In the source considered in Subsection \ref{transitionevidence}, which possesses a pumping speed of $3000\ \mathrm{L\,s^{-1}}$ and a nozzle-skimmer distance of $\sim10\ \mathrm{mm}$, skimmer effects are visible near $150\ \mathrm{K}$ but only dominate below $\sim50\ \mathrm{K}$. We have also investigated interference in a series of hypothetical source chambers, assuming low nozzle temperatures, to develop a set of design recommendations. To summarize, we find that (i) surfaces perpendicular to the beam are surprisingly problematic, (ii) narrow skimmers are not necessarily superior to broader ones and (iii) the theory can be used to optimize skimmer designs.

The benefits of the theory are twofold. First, source chamber interference can now be accounted for straightforwardly when estimating centerline intensities. Even if one suspects that there will be negligible attenuation, it is now easy to check, via the condition $\eta Q\ll1$. Second, the theory allows for the optimization of skimmers and source chambers. Estimates can now be made of the interference due to proposed designs, and they may be rejected before more effort is devoted to developing them. We anticipate that improvements to existing skimmers may be made by combining our theory with further experimental work on internal/external shock waves.


\section{Acknowledgements}
The work was supported by EPSRC grant EP/R008272/1 and Innovate UK/Ionoptika Ltd. through Knowledge Transfer Partnership 10000925. The work was carried out in part at CORDE, the Collaborative R\&D Environment established to provide access to physics-related facilities at the Cavendish Laboratory, University of Cambridge, and the EPSRC award EP / T00634X / 1. SML acknowledges support from EPSRC grant EP/X525686/1. The authors appreciate helpful discussion and comments from Boyao Liu, Paul Dastoor and William Allison.

\section*{Author Declarations}

\subsection*{Conflicts of Interest}
The authors have no conflicts to disclose.

\subsection*{Author Contributions}
\textbf{Jack Kelsall:} methodology (lead), formal analysis (lead), review and editing (support), writing - original draft preparation (equal). \textbf{Aleksandar Radi\'{c}:} writing - original draft preparation (equal), writing - review and editing (lead). \textbf{John Ellis:} conceptualization (equal), supervision (supporting). \textbf{David J. Ward:} conceptualization (equal), supervision (lead). \textbf{Andrew P. Jardine:} supervision (supporting), funding acquisition (lead).

\subsection*{Data Availability}
All code used in this study will be available at the University of Cambridge repository (Apollo) upon publication at \url{https://doi.org/10.17863/CAM.115338}.

\appendix

\section{Complete Derivation of Cross Section}
\label{crosssectiondev}

\renewcommand{\figurename}{Fig.}
\renewcommand{\thefigure}{\arabic{figure}}

Here we derive the cross section describing the interaction between a beam atom and an interfering particle. Much of the derivation here builds on earlier work by Hedgeland et al. \cite{anomalous}, although the final result in the present case is more general.

Consider a particle of the beam travelling along the $z$-axis with speed $v$. Let the speed of a residual gas molecule be $u$ and the velocity's direction angle to the negative $z$-direction be $\theta$. The situation is illustrated in the first panel of Figure \ref{frames}. The center of mass (CoM) frame is also depicted there, alongside the instantaneous rest frame (IRF) of the beam atom. In the latter, the speed of the residual atom is $w=\sqrt{u^2+v^2+2uv\cos\theta}$ and the angle $\delta$ is given by $\tan\delta=u\sin\theta/\left(v+u\cos\theta\right)$.

The critical impact parameter for interference with the beam is determined by small-angle scattering. As a result, in the limiting case that defines the cross section, the atoms will barely be deflected from their initial straight-line paths due to their mutual interaction. Let the separation of the two atoms at an instant be $r=2r_c$ and the minimum value (i.e. the distance of closest approach) be $d$. The trajectory $r_c=r_c\left(\psi\right)$ can therefore be approximated by

\begin{equation}
	\frac{d}{2}=r_c\sin\psi, 
	\label{solution}
\end{equation}

\noindent in the CoM frame. Using that lowest-order solution, it is possible to calculate the perpendicular impulse $I_\perp=\int F_\perp \mathrm{d}t=\int F\sin\psi \mathrm{d}t$, assuming that a van der Waals (vdW) potential,

\begin{equation}
	V\left(r\right)=-\frac{C_6}{r^6},
	\label{vdWpot}
\end{equation}

\noindent is a suitable model. The central force is given by $F=-\mathrm{d}V/\mathrm{d}r$. $I_\perp$ is the only non-zero component of the impulse, by symmetry. Both experimental \cite{scoles_beam_methods} and theoretical \cite{c6valuescomput} values of the coefficient $C_6$ exist; the DFT value of $C_6=1.46\ \mathrm{a.u.}$ is used in all computations in the main text. Here, $1\ \mathrm{a.u.} $ (atomic unit) is equal to $E_h/a_0^6=\hbar^2/m_ea_0^8$ for Hartree energy $E_h$, electronic mass $m_e$ and Bohr radius $a_0$.

The final expression needed to evaluate the integral for $I_\perp$ comes from conservation of the angular momentum, $L$, of one particle about the CoM, $O$,

\begin{equation}
	L=\frac{mwd}{4}=mr_c^2\dot{\psi},
	\label{angmomentum}
\end{equation}

\noindent so $\dot{\psi}=wd/4r_c^2$ and $\mathrm{d}t=\mathrm{d}\psi/\dot{\psi}=4r_c^2\mathrm{d}\psi/wd$. $m$ is the mass of a gas atom. Hence, the impulse is

\begin{equation}
	I_\perp=\frac{3C_6}{16wd}\int_{0}^{\pi}{\frac{\sin{\psi}}{r_c^5}\mathrm{d}\psi}=\frac{{15\pi C}_6}{8wd^6}.
	\label{finalI0}
\end{equation}

\begin{figure}[h]
	\includegraphics[width=0.5\textwidth]{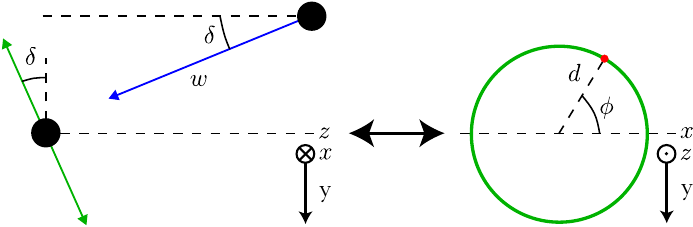}
	\centering
	\caption{A sketch showing the circle of possible impulses for a given impact parameter, $d$, and hence beam-perpendicular impulse $I_\perp$. The trajectory of backscattered or background gas atom, indicated by a red dot on the edge of the circle, is parameterized by the angle $\varphi$, which is measured with respect to the $x$-axis. The total impulse, integrated over the entire trajectory, is perpendicular to the path of the interfering atom.}
	\label{impulse_diagram}
\end{figure}

Now consider an atom traveling towards the beam particle in its IRF at the angle $\delta$. The impulse is invariant under the Galilean transformation that relates the second and third frames of Figure \ref{frames}, so \eqref{finalI0} applies without modification. Let the angle $\varphi$ parameterize the path of a particular atom, as illustrated by Figure \ref{impulse_diagram}, so that the components of the impulse acting on the beam atom are

\begin{equation}
	\bm{I}=I_\perp
	\begin{pmatrix}
		\cos\varphi\\
		\sin\varphi\cos\delta\\
		\sin\varphi\sin\delta
	\end{pmatrix}
	,
	\label{impulsecomponents}
\end{equation}

\noindent with the $x$-axis into the page and $z$. Only the first two components result in a deflection of the beam atom away from the axis, so the magnitude of the relevant impulse is

\begin{equation}
	I=I_\perp\sqrt{\cos^2\varphi+\sin^2\varphi\cos^2\delta}.
	\label{deflectingimpulse}
\end{equation}

Let $\Omega$ be the angle subtended by the radius of the limiting aperture from the nozzle. For an atom to be deflected such that it fails to reach the detector, its trajectory must be perturbed by at least $\Omega$. In the limiting case, $I=mv\Omega$, via the small-angle approximation. Let $d_0$ be the corresponding critical impact parameter. Inserting expressions and rearranging for $d_0$,

\begin{equation}
	d_0=\left(\frac{15\pi C_6}{8mvw\Omega}\right)^{1/6}\left(\cos^2\varphi+\sin^2\varphi\cos^2\delta\right)^{1/12}.
	\label{criticald0}
\end{equation}

$\Omega$ is unlikely to be the angle of the skimmer aperture viewed from the nozzle, which is typically large compared to e.g. a pre-detector aperture. Furthermore, given that $d<d_0$ leads to deflection, $d_0$ defines a cross section, $\sigma=\left(1/2\right)\int_{-\pi}^\pi d_0^2\mathrm{d}\varphi$. Using \eqref{criticald0},

\begin{equation}
	\sigma=\left(\frac{15\pi C_6}{8mvw\Omega}\right)^{1/3}\int_{-\pi/2}^{\pi/2}\left(\cos^2\varphi+\sin^2\varphi\cos^2\delta\right)^{1/6}\mathrm{d}\varphi.
	\label{crosssectionrep}
\end{equation}

The integral in \eqref{crosssectionrep} cannot be evaluated analytically. The result is a generalization of that in Hedgeland et al. \cite{anomalous} to larger angles $\delta$; both expressions are identical to second-order. $\sigma$ is much larger than a typical hard-sphere cross section because weak interactions are sufficient to cause deflections that prevent transmission to the detector in long-beam systems.

\section{Extension of backscattering formulae to a continuous profile function}
\label{fappendix}

\renewcommand{\figurename}{Fig.}
\renewcommand{\thefigure}{\arabic{figure}}

Equation \eqref{etas} for the backscattering attenuation coefficient can be modified for a continuous profile function, $f(x)$, describing the skimmer and source chamber. Here, $x$ is the horizontal distance of a point from the tip of the skimmer. The integral over $r$ is thus replaced by one over $x$.

An infinitesimal element of the profile has length $\mathrm{d}x$ and height $\mathrm{d}f$. Its angle to the $x$-axis is therefore given by

\begin{equation}
	\tan{\gamma=f'(x)}.
	\label{tangammaf}
\end{equation}

\noindent The effective nozzle-tip distance for that element is

\begin{equation}
	d_E=x+d-f\left(x\right)/f'\left(x\right),
	\label{def}
\end{equation}

\noindent The distance of a point $P$ from the effective tip, at the distance $d_E$, is

\begin{equation}
	r=f(x)\sqrt{1+\left[f'\left(x\right)\right]^{-2}}.
	\label{rf}
\end{equation}

\noindent The total length of the element is

\begin{equation}
	\mathrm{d}r=\sqrt{1+\left[f'\left(x\right)\right]^2}\mathrm{d}x.
	\label{drf}
\end{equation}

In Appendix \ref{backscat_app}, we use this result to evaluate the performance of a parabolic skimmer profile as a function of the curvature parameter, $\epsilon$, defined by $f(x)=\epsilon x^2$.

\section{Beam attenuation due to backscattered particles - exact and approximate solutions}
\label{backscat_app}

\renewcommand{\figurename}{Fig.}
\renewcommand{\thefigure}{\arabic{figure}}

\begin{figure}[H]
    \centering
    \includegraphics[width=\linewidth]{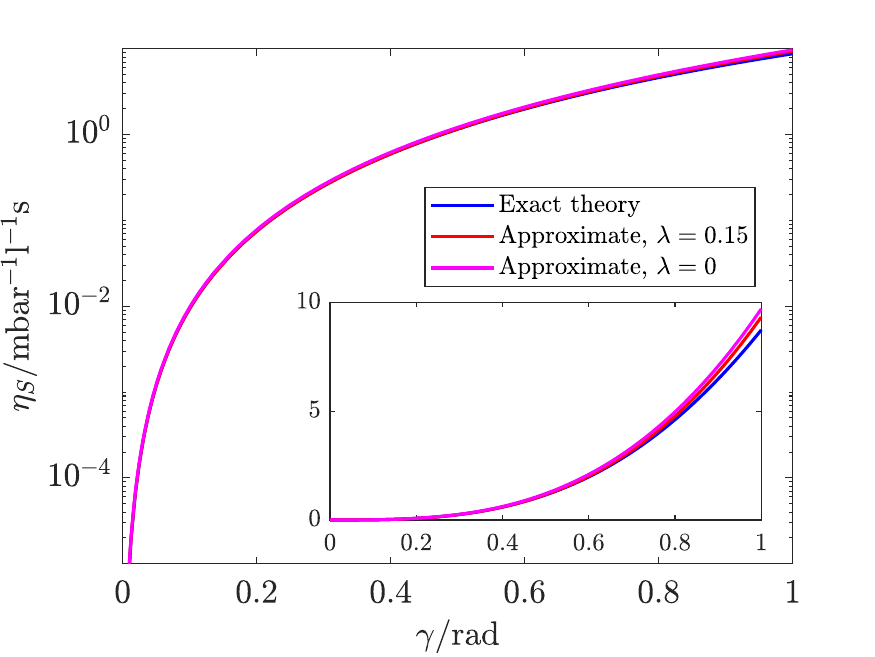}
    \caption{A plot of the backscattering attenuation coefficient, $\eta_S$, as a function of the half angle, $\gamma$, defining a conical skimmer. The ramped length and nozzle-skimmer distance are $a=d_B=15\ \mathrm{mm}$. Both the exact and approximate expressions for $\eta_S$ have been evaluated and plotted, corresponding to Equations \eqref{etas} and \eqref{etasconical_maintext}, respectively. The same graph with a linear scale is included in the inset.}
    \label{backscattered_approx}
\end{figure}

We first consider a conical skimmer, with an approximate expression for $\eta_S$ given by \eqref{etasconical_maintext}. We compare the numerical solution to the exact integral, i.e. \eqref{etas}, for a range of skimmer angles. In Figure \ref{backscattered_approx}, we evaluate (i) the exact numerical solution to \eqref{etas}, (ii) the approximate result with $\lambda=0.15$, and (iii) the approximate result with $\lambda=0$, all as a function of the half angle at the skimmer tip, $\gamma$.

From Figure \ref{backscattered_approx}, we can see there is excellent agreement between the exact and approximate attenuation coefficients due to backscattered particles, even as $\gamma\rightarrow1$. A significant finding from \eqref{etasconical_maintext} is that $\eta_S\propto\gamma^3$.

\begin{figure}[h]
	\includegraphics[width=\linewidth]{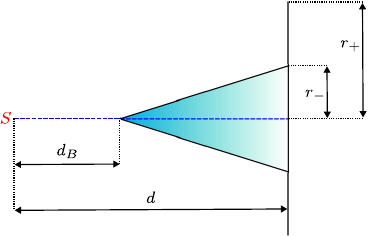}
	\centering
	\caption{A skimmer mount. The skimmer, in blue, obscures the part of the beam nearest the mount, so although the nozzle-mount distance is $d$, one must integrate from $z=d-d_B$ to $d$ when calculating $\eta_S$. For $r_+\ll d-d_B$, it is straightforward to find an approximate expression for $\eta_S$.}
	\label{backplate}
\end{figure}

Finding an expression for $\eta_S$ for a general plate (i.e. an element with $\gamma=\pi/2$) is more difficult than doing so for a conical skimmer because a wide range of scattering angles contribute to the attenuation when the element is perpendicular to the beam. Therefore, for a plate, it is most straightforward to numerically integrate \eqref{etas}. However, an approximate analytic result can be found for the special case of a skimmer mount, positioned such that the skimmer limits the range of scattering angles. The geometry is illustrated by Figure \ref{backplate}.

Consider a small mount, extending from $r=r_-$ to $r_+$ with $r_-,r_+\ll d_B$. Let $d$ be the distance between the nozzle and the mount, for consistency with Figure \ref{fig:conical_schematic}. Assuming that all angles (except $\gamma=\pi/2$) are small leads to

\begin{equation}
	\eta_S\approx\frac{p+1}{4k_BT_0}\left(\frac{15\pi C_6}{mv_0^4\Omega}\right)^{1/3}\frac{d_B\left(r_+^2-r_-^2\right)}{d^3\left(d-d_B\right)}I_u.
	\label{etasplate}
\end{equation}

\begin{figure}[h]
	\includegraphics[width=\linewidth]{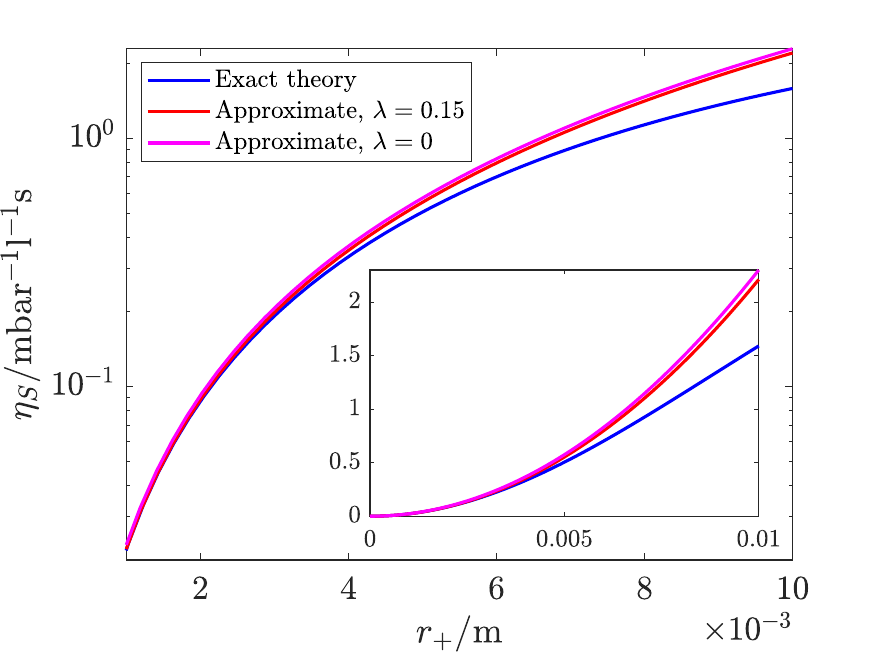}
	\centering
	\caption{A plot of the backscattering attenuation coefficient, $\eta_S$, as a function of outer radius, $r_+$, for a skimmer mount. Figure \ref{backplate} defines the important quantities; we take $d=d_B$ and $r_-=0$. Both the exact and approximate expressions have been evaluated and plotted, corresponding to Equations \eqref{etas} and \eqref{etasplate}, respectively. The same graph with a linear scale is included in the inset.}
	\label{plate_approximate}
\end{figure}

\noindent To further simplify the above, $\lambda=0$ can be inserted. In Figure \ref{plate_approximate}, we plot (i) the exact result, (ii) the approximate result with $\lambda=0.15$ and (iii) the approximate result with $\lambda=0$. We use $d_B=15\ \mathrm{mm}$, $d=30\ \mathrm{mm}$ and $r_-=0$. The contribution from the skimmer is neglected.

\begin{figure}[h]
	\includegraphics[width=\linewidth]{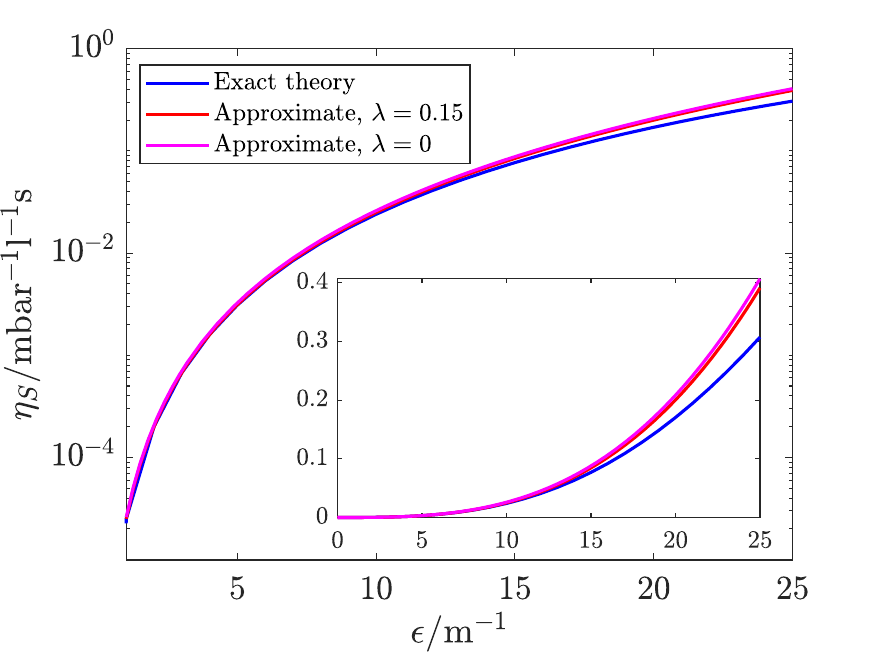}
	\centering
	\caption{A plot of the backscattering attenuation coefficient, $\eta_S$, as a function of the curvature parameter, $\epsilon$, describing a parabolic skimmer with a profile given by $f\left(x\right)=\epsilon x^2$. The nozzle-skimmer distance and skimmer length are $d_B=x_2=15\ \mathrm{mm}$, and we take $x_1=0$. Both the exact and approximate expressions have been evaluated, corresponding to Equations \eqref{etas} and \eqref{etasparabolic}. The same graph with a linear scale is plotted in the inset.}
	\label{parabolic_approximate}
\end{figure}

To determine $\eta_S$ due to a narrow parabolic skimmer, we follow the approach outlined in Appendix \ref{fappendix}, i.e. we describe the profile in terms of the horizontal distance, $x$, from the tip. The skimmer may then be described by a profile function $f\left(x\right)=\epsilon x^2$. As long as $\max\left(f\right)\ll d_B$, the attenuation coefficient evaluates to

\begin{multline}
	\eta_S\approx\frac{\left(p+1\right)d_B\epsilon^3}{2k_BT_0}\left(\frac{15\pi C_6}{mv_0^4\Omega}\right)^{1/3}I_u\\
	\times\left\{\left(\frac{d_B}{x+d_B}\right)^2\left[x+\frac{d_B^3}{8\left(x+d_B\right)^2}\right]\right.\\
	\left.\left.-\frac{3}{2}d_B\ln{\left(x+d_B\right)}+x\right\}\right|_{x_1}^{x_2},
	\label{etasparabolic}
\end{multline}

\noindent where $x_1$ and $x_2$ define the extent of the skimmer. By altering the limits, one can approximate different curves (e.g. circular arcs) by parabolas to calculate their attenuation coefficients. In Figure \ref{parabolic_approximate}, we plot (i) the exact result from Equation \eqref{etas}, (ii) the approximate result with $\lambda=0.15$ and (iii) the approximate result with $\lambda=0$. We take $x_1=0$ and $x_2=d_B=15\ \mathrm{mm}$.

\section{Beam attenuation due to background particles - exact and approximate solutions}
\label{background_app}

\renewcommand{\figurename}{Fig.}
\renewcommand{\thefigure}{\arabic{figure}}

Here we compare the exact integral for the attenuation coefficient due to the diffuse background, $\eta_B$ of \eqref{etab}, and its approximate analytic form, \eqref{etabapprox}. In the following calculations, we take the nozzle-skimmer distance to be $d_B=\SI{15}{mm}$ and consider a \textsuperscript{4}He beam with a representative value of $\Omega=\SI{1.366e-3}{\radian}$. The chamber temperature is set to $T_C=293\ \mathrm{K}$.

\begin{figure}[h]
	\includegraphics[width=\linewidth]{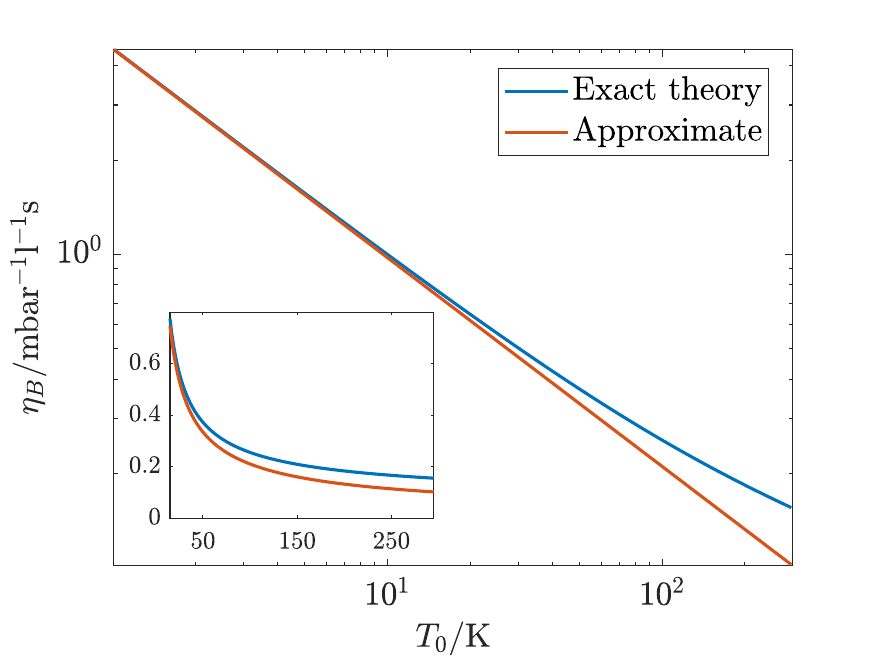}
	\centering
	\caption{A plot of the attenuation coefficient due to background gas in the chamber, $\eta_B$, as a function of nozzle temperature, $T_0$, for $S=2500\ \mathrm{ls^{-1}}$. Both the exact and approximate results have been evaluated and plotted, corresponding to \eqref{etab} and \eqref{etabapprox}, respectively.}
	\label{background_approximate}
\end{figure}

$\eta_B$ is independent of the geometry of the source, making it difficult to approximate in general. However, for nozzle temperatures $T_0\ll T_C$, an analytic expression can be derived by noting that most of the interfering molecules have speeds, $u$, much larger than the beam velocity, $v_0$. As a result, the relative speed, $w$, satisfies $w\approx u$ in most scattering events. Evaluatingt the resulting separable integrals leads to \eqref{etabapprox}. We plot the exact and approximate expressions for $\eta_B$ in Figure \ref{background_approximate}, using $S=2500\ \mathrm{ls^{-1}}$. However, a different pumping speed simply leads to a rescaling of all values of the coefficient, because $\eta_B\propto1/S$.

Figure \ref{background_approximate} shows excellent agreement for cryogenic nozzle temperatures, but the fractional error rises as $T_0\rightarrow T_C$, where $T_C=\SI{293}{\kelvin}$ is the chamber temperature.

\newpage
\section{Skimmer Geometries}
\label{skimmer_geometries}

\renewcommand{\figurename}{Fig.}
\renewcommand{\thefigure}{\arabic{figure}}

\begin{figure}[h]
\centering
\includegraphics[width=\linewidth]{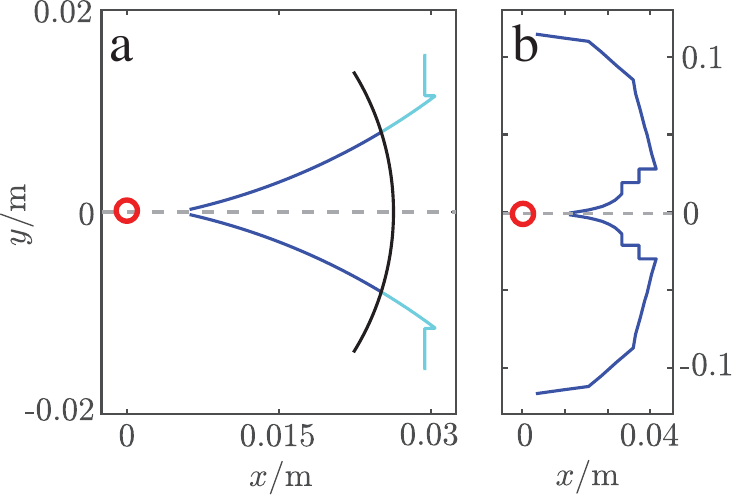}
\caption{Schematic cross sections of the two skimmer geometries used to experimentally investigate beam attenuation, relating to data presented in Figure 3. The source nozzle is marked by a red circle, the beam axis is the horizontal dashed line, the Mach disc is colored black and the skimmer itself is blue/cyan. The two skimmer colors represent regions that lie within (blue) and outside the Mach disc (cyan). Source 1 uses the geometry shown in Panel (a), while Source 3 corresponds to Panel (b), in which more of the surrounding chamber is relevant. The position of the Mach disc in Panel (a) was calculated using $T_0=21\ \mathrm{K}$.}
\label{skimmer_geometries_figure}
\end{figure}

Figure \ref{skimmer_geometries_figure} shows schematic diagrams of the skimmer geometries used in the experimental validation of our theory, corresponding to the data in Figure \ref{fig:big_figure}. Only the region around and near the skimmer is relevant in Source 1, corresponding to Panel (a), while Source 3 possesses the more complex geometry in Panel (b). The attenuation in Source 2 is approximately independent of the geometry because it is dominated by the background contribution.

\clearpage
\bibliography{bibliographypaper}

\begin{thebibliography}{47}%
\makeatletter
\providecommand \@ifxundefined [1]{%
 \@ifx{#1\undefined}
}%
\providecommand \@ifnum [1]{%
 \ifnum #1\expandafter \@firstoftwo
 \else \expandafter \@secondoftwo
 \fi
}%
\providecommand \@ifx [1]{%
 \ifx #1\expandafter \@firstoftwo
 \else \expandafter \@secondoftwo
 \fi
}%
\providecommand \natexlab [1]{#1}%
\providecommand \enquote  [1]{``#1''}%
\providecommand \bibnamefont  [1]{#1}%
\providecommand \bibfnamefont [1]{#1}%
\providecommand \citenamefont [1]{#1}%
\providecommand \href@noop [0]{\@secondoftwo}%
\providecommand \href [0]{\begingroup \@sanitize@url \@href}%
\providecommand \@href[1]{\@@startlink{#1}\@@href}%
\providecommand \@@href[1]{\endgroup#1\@@endlink}%
\providecommand \@sanitize@url [0]{\catcode `\\12\catcode `\$12\catcode `\&12\catcode `\#12\catcode `\^12\catcode `\_12\catcode `\%12\relax}%
\providecommand \@@startlink[1]{}%
\providecommand \@@endlink[0]{}%
\providecommand \url  [0]{\begingroup\@sanitize@url \@url }%
\providecommand \@url [1]{\endgroup\@href {#1}{\urlprefix }}%
\providecommand \urlprefix  [0]{URL }%
\providecommand \Eprint [0]{\href }%
\providecommand \doibase [0]{https://doi.org/}%
\providecommand \selectlanguage [0]{\@gobble}%
\providecommand \bibinfo  [0]{\@secondoftwo}%
\providecommand \bibfield  [0]{\@secondoftwo}%
\providecommand \translation [1]{[#1]}%
\providecommand \BibitemOpen [0]{}%
\providecommand \bibitemStop [0]{}%
\providecommand \bibitemNoStop [0]{.\EOS\space}%
\providecommand \EOS [0]{\spacefactor3000\relax}%
\providecommand \BibitemShut  [1]{\csname bibitem#1\endcsname}%
\let\auto@bib@innerbib\@empty
\bibitem [{\citenamefont {Kirkpatrick}\ \emph {et~al.}(2016)\citenamefont {Kirkpatrick}, \citenamefont {Walsh}, \citenamefont {Svrluga},\ and\ \citenamefont {Khoury}}]{anab}%
  \BibitemOpen
  \bibfield  {author} {\bibinfo {author} {\bibfnamefont {S.}~\bibnamefont {Kirkpatrick}}, \bibinfo {author} {\bibfnamefont {M.}~\bibnamefont {Walsh}}, \bibinfo {author} {\bibfnamefont {R.}~\bibnamefont {Svrluga}},\ and\ \bibinfo {author} {\bibfnamefont {J.}~\bibnamefont {Khoury}},\ }\bibfield  {title} {\bibinfo {title} {Accelerated neutral atom beam ({ANAB}) technology for nanoscale surface processing},\ }in\ \href {https://doi.org/10.1109/NANO.2016.7751397} {\emph {\bibinfo {booktitle} {2016 IEEE 16th International Conference on Nanotechnology (IEEE-NANO)}}}\ (\bibinfo {year} {2016})\ p.\ \bibinfo {pages} {710}\BibitemShut {NoStop}%
\bibitem [{\citenamefont {Allred}\ \emph {et~al.}(2010)\citenamefont {Allred}, \citenamefont {Reeves}, \citenamefont {Corder},\ and\ \citenamefont {Metcalf}}]{Allred2010}%
  \BibitemOpen
  \bibfield  {author} {\bibinfo {author} {\bibfnamefont {C.~S.}\ \bibnamefont {Allred}}, \bibinfo {author} {\bibfnamefont {J.}~\bibnamefont {Reeves}}, \bibinfo {author} {\bibfnamefont {C.}~\bibnamefont {Corder}},\ and\ \bibinfo {author} {\bibfnamefont {H.}~\bibnamefont {Metcalf}},\ }\bibfield  {title} {\bibinfo {title} {Atom lithography with metastable helium},\ }\bibfield  {journal} {\bibinfo  {journal} {Journal of Applied Physics}\ }\textbf {\bibinfo {volume} {107}},\ \href {https://doi.org/10.1063/1.3295903} {10.1063/1.3295903} (\bibinfo {year} {2010})\BibitemShut {NoStop}%
\bibitem [{\citenamefont {Berggren}\ \emph {et~al.}(1995)\citenamefont {Berggren}, \citenamefont {Bard}, \citenamefont {Wilbur}, \citenamefont {Gillaspy}, \citenamefont {Helg}, \citenamefont {McClelland}, \citenamefont {Rolston}, \citenamefont {Phillips}, \citenamefont {Prentiss},\ and\ \citenamefont {Whitesides}}]{Berggren_1995}%
  \BibitemOpen
  \bibfield  {author} {\bibinfo {author} {\bibfnamefont {K.~K.}\ \bibnamefont {Berggren}}, \bibinfo {author} {\bibfnamefont {A.}~\bibnamefont {Bard}}, \bibinfo {author} {\bibfnamefont {J.~L.}\ \bibnamefont {Wilbur}}, \bibinfo {author} {\bibfnamefont {J.~D.}\ \bibnamefont {Gillaspy}}, \bibinfo {author} {\bibfnamefont {A.~G.}\ \bibnamefont {Helg}}, \bibinfo {author} {\bibfnamefont {J.~J.}\ \bibnamefont {McClelland}}, \bibinfo {author} {\bibfnamefont {S.~L.}\ \bibnamefont {Rolston}}, \bibinfo {author} {\bibfnamefont {W.~D.}\ \bibnamefont {Phillips}}, \bibinfo {author} {\bibfnamefont {M.}~\bibnamefont {Prentiss}},\ and\ \bibinfo {author} {\bibfnamefont {G.~M.}\ \bibnamefont {Whitesides}},\ }\bibfield  {title} {\bibinfo {title} {Microlithography by using neutral metastable atoms and self-assembled monolayers},\ }\href {https://doi.org/10.1126/science.7652572} {\bibfield  {journal} {\bibinfo  {journal} {Science}\ }\textbf {\bibinfo {volume} {269}},\ \bibinfo {pages} {1255–1257} (\bibinfo {year}
  {1995})}\BibitemShut {NoStop}%
\bibitem [{\citenamefont {von Jeinsen}\ \emph {et~al.}(2023)\citenamefont {von Jeinsen}, \citenamefont {Lambrick}, \citenamefont {Bergin}, \citenamefont {Radić}, \citenamefont {Liu}, \citenamefont {Seremet}, \citenamefont {Jardine},\ and\ \citenamefont {Ward}}]{von_jeinsen_2d_2023}%
  \BibitemOpen
  \bibfield  {author} {\bibinfo {author} {\bibfnamefont {N.~A.}\ \bibnamefont {von Jeinsen}}, \bibinfo {author} {\bibfnamefont {S.~M.}\ \bibnamefont {Lambrick}}, \bibinfo {author} {\bibfnamefont {M.}~\bibnamefont {Bergin}}, \bibinfo {author} {\bibfnamefont {A.}~\bibnamefont {Radić}}, \bibinfo {author} {\bibfnamefont {B.}~\bibnamefont {Liu}}, \bibinfo {author} {\bibfnamefont {D.}~\bibnamefont {Seremet}}, \bibinfo {author} {\bibfnamefont {A.~P.}\ \bibnamefont {Jardine}},\ and\ \bibinfo {author} {\bibfnamefont {D.~J.}\ \bibnamefont {Ward}},\ }\bibfield  {title} {\bibinfo {title} {{2D} {Helium} {Atom} {Diffraction} from a {Microscopic} {Spot}},\ }\href {https://doi.org/10.1103/PhysRevLett.131.236202} {\bibfield  {journal} {\bibinfo  {journal} {Physical Review Letters}\ }\textbf {\bibinfo {volume} {131}},\ \bibinfo {pages} {236202} (\bibinfo {year} {2023})},\ \bibinfo {note} {publisher: American Physical Society}\BibitemShut {NoStop}%
\bibitem [{\citenamefont {Radic}\ \emph {et~al.}(2024)\citenamefont {Radic}, \citenamefont {von Jeinsen}, \citenamefont {Wang}, \citenamefont {Zhu}, \citenamefont {Sami}, \citenamefont {Perez}, \citenamefont {Ward}, \citenamefont {Jardine}, \citenamefont {Chhowalla},\ and\ \citenamefont {Lambrick}}]{defects_radic}%
  \BibitemOpen
  \bibfield  {author} {\bibinfo {author} {\bibfnamefont {A.}~\bibnamefont {Radic}}, \bibinfo {author} {\bibfnamefont {N.}~\bibnamefont {von Jeinsen}}, \bibinfo {author} {\bibfnamefont {K.}~\bibnamefont {Wang}}, \bibinfo {author} {\bibfnamefont {Y.}~\bibnamefont {Zhu}}, \bibinfo {author} {\bibfnamefont {I.}~\bibnamefont {Sami}}, \bibinfo {author} {\bibfnamefont {V.}~\bibnamefont {Perez}}, \bibinfo {author} {\bibfnamefont {D.}~\bibnamefont {Ward}}, \bibinfo {author} {\bibfnamefont {A.}~\bibnamefont {Jardine}}, \bibinfo {author} {\bibfnamefont {M.}~\bibnamefont {Chhowalla}},\ and\ \bibinfo {author} {\bibfnamefont {S.}~\bibnamefont {Lambrick}},\ }\bibfield  {title} {\bibinfo {title} {Defect density quantification in monolayer {MoS2} using helium atom micro-diffraction},\ }\href@noop {} {\bibfield  {journal} {\bibinfo  {journal} {arXiv}\ } (\bibinfo {year} {2024})}\BibitemShut {NoStop}%
\bibitem [{\citenamefont {von Jeinsen}\ \emph {et~al.}(2024)\citenamefont {von Jeinsen}, \citenamefont {Radic}, \citenamefont {Wang}, \citenamefont {Zhao}, \citenamefont {Perez}, \citenamefont {Zhu}, \citenamefont {Chhowalla}, \citenamefont {Jardine}, \citenamefont {Ward},\ and\ \citenamefont {Lambrick}}]{2d_methods}%
  \BibitemOpen
  \bibfield  {author} {\bibinfo {author} {\bibfnamefont {N.}~\bibnamefont {von Jeinsen}}, \bibinfo {author} {\bibfnamefont {A.}~\bibnamefont {Radic}}, \bibinfo {author} {\bibfnamefont {K.}~\bibnamefont {Wang}}, \bibinfo {author} {\bibfnamefont {C.}~\bibnamefont {Zhao}}, \bibinfo {author} {\bibfnamefont {V.}~\bibnamefont {Perez}}, \bibinfo {author} {\bibfnamefont {Y.}~\bibnamefont {Zhu}}, \bibinfo {author} {\bibfnamefont {M.}~\bibnamefont {Chhowalla}}, \bibinfo {author} {\bibfnamefont {A.}~\bibnamefont {Jardine}}, \bibinfo {author} {\bibfnamefont {D.}~\bibnamefont {Ward}},\ and\ \bibinfo {author} {\bibfnamefont {S.}~\bibnamefont {Lambrick}},\ }\bibfield  {title} {\bibinfo {title} {Helium atom micro-diffraction as a characterisation tool for {2D} materials},\ }\href@noop {} {\bibfield  {journal} {\bibinfo  {journal} {arXiv}\ } (\bibinfo {year} {2024})}\BibitemShut {NoStop}%
\bibitem [{\citenamefont {Flatabø}\ \emph {et~al.}(2024)\citenamefont {Flatabø}, \citenamefont {Eder}, \citenamefont {Reisinger}, \citenamefont {Bracco}, \citenamefont {Baltzer}, \citenamefont {Samelin},\ and\ \citenamefont {Holst}}]{FLATABO2024113961}%
  \BibitemOpen
  \bibfield  {author} {\bibinfo {author} {\bibfnamefont {R.}~\bibnamefont {Flatabø}}, \bibinfo {author} {\bibfnamefont {S.~D.}\ \bibnamefont {Eder}}, \bibinfo {author} {\bibfnamefont {T.}~\bibnamefont {Reisinger}}, \bibinfo {author} {\bibfnamefont {G.}~\bibnamefont {Bracco}}, \bibinfo {author} {\bibfnamefont {P.}~\bibnamefont {Baltzer}}, \bibinfo {author} {\bibfnamefont {B.}~\bibnamefont {Samelin}},\ and\ \bibinfo {author} {\bibnamefont {Holst}},\ }\bibfield  {title} {\bibinfo {title} {Reflection imaging with a helium zone plate microscope},\ }\href {https://doi.org/https://doi.org/10.1016/j.ultramic.2024.113961} {\bibfield  {journal} {\bibinfo  {journal} {Ultramicroscopy}\ }\textbf {\bibinfo {volume} {261}},\ \bibinfo {pages} {113961} (\bibinfo {year} {2024})}\BibitemShut {NoStop}%
\bibitem [{\citenamefont {Palau}\ \emph {et~al.}(2023)\citenamefont {Palau}, \citenamefont {Eder}, \citenamefont {Bracco},\ and\ \citenamefont {Holst}}]{Palau2023}%
  \BibitemOpen
  \bibfield  {author} {\bibinfo {author} {\bibfnamefont {A.~S.}\ \bibnamefont {Palau}}, \bibinfo {author} {\bibfnamefont {S.~D.}\ \bibnamefont {Eder}}, \bibinfo {author} {\bibfnamefont {G.}~\bibnamefont {Bracco}},\ and\ \bibinfo {author} {\bibfnamefont {B.}~\bibnamefont {Holst}},\ }\bibfield  {title} {\bibinfo {title} {Neutral helium atom microscopy},\ }\href {https://doi.org/10.1016/j.ultramic.2023.113753} {\bibfield  {journal} {\bibinfo  {journal} {Ultramicroscopy}\ }\textbf {\bibinfo {volume} {251}},\ \bibinfo {pages} {113753} (\bibinfo {year} {2023})}\BibitemShut {NoStop}%
\bibitem [{\citenamefont {Radić}\ \emph {et~al.}(2024)\citenamefont {Radić}, \citenamefont {Lambrick}, \citenamefont {von Jeinsen}, \citenamefont {Jardine},\ and\ \citenamefont {Ward}}]{radic_3d_2024}%
  \BibitemOpen
  \bibfield  {author} {\bibinfo {author} {\bibfnamefont {A.}~\bibnamefont {Radić}}, \bibinfo {author} {\bibfnamefont {S.~M.}\ \bibnamefont {Lambrick}}, \bibinfo {author} {\bibfnamefont {N.~A.}\ \bibnamefont {von Jeinsen}}, \bibinfo {author} {\bibfnamefont {A.~P.}\ \bibnamefont {Jardine}},\ and\ \bibinfo {author} {\bibfnamefont {D.~J.}\ \bibnamefont {Ward}},\ }\bibfield  {title} {\bibinfo {title} {{3D} surface profilometry using neutral helium atoms},\ }\href {https://doi.org/10.1063/5.0206374} {\bibfield  {journal} {\bibinfo  {journal} {Applied Physics Letters}\ }\textbf {\bibinfo {volume} {124}},\ \bibinfo {pages} {204101} (\bibinfo {year} {2024})}\BibitemShut {NoStop}%
\bibitem [{\citenamefont {Lambrick}\ \emph {et~al.}(2022{\natexlab{a}})\citenamefont {Lambrick}, \citenamefont {Bergin}, \citenamefont {Ward}, \citenamefont {Barr}, \citenamefont {Fahy}, \citenamefont {Myles}, \citenamefont {Radić}, \citenamefont {Dastoor}, \citenamefont {Ellis},\ and\ \citenamefont {Jardine}}]{LambrickDiffuse2022}%
  \BibitemOpen
  \bibfield  {author} {\bibinfo {author} {\bibfnamefont {S.~M.}\ \bibnamefont {Lambrick}}, \bibinfo {author} {\bibfnamefont {M.}~\bibnamefont {Bergin}}, \bibinfo {author} {\bibfnamefont {D.~J.}\ \bibnamefont {Ward}}, \bibinfo {author} {\bibfnamefont {M.}~\bibnamefont {Barr}}, \bibinfo {author} {\bibfnamefont {A.}~\bibnamefont {Fahy}}, \bibinfo {author} {\bibfnamefont {T.}~\bibnamefont {Myles}}, \bibinfo {author} {\bibfnamefont {A.}~\bibnamefont {Radić}}, \bibinfo {author} {\bibfnamefont {P.~C.}\ \bibnamefont {Dastoor}}, \bibinfo {author} {\bibfnamefont {J.}~\bibnamefont {Ellis}},\ and\ \bibinfo {author} {\bibfnamefont {A.~P.}\ \bibnamefont {Jardine}},\ }\bibfield  {title} {\bibinfo {title} {Observation of diffuse scattering in scanning helium microscopy},\ }\href {https://doi.org/10.1039/D2CP01951E} {\bibfield  {journal} {\bibinfo  {journal} {Phys. Chem. Chem. Phys.}\ }\textbf {\bibinfo {volume} {24}},\ \bibinfo {pages} {26539} (\bibinfo {year} {2022}{\natexlab{a}})}\BibitemShut {NoStop}%
\bibitem [{\citenamefont {Jankunas}\ and\ \citenamefont {Osterwalder}(2015)}]{coldmolbeams}%
  \BibitemOpen
  \bibfield  {author} {\bibinfo {author} {\bibfnamefont {J.}~\bibnamefont {Jankunas}}\ and\ \bibinfo {author} {\bibfnamefont {A.}~\bibnamefont {Osterwalder}},\ }\bibfield  {title} {\bibinfo {title} {Cold and controlled molecular beams: Production and applications},\ }\href {https://doi.org/10.1146/annurev-physchem-040214-121307} {\bibfield  {journal} {\bibinfo  {journal} {Annu. Rev. Phys. Chem.}\ }\textbf {\bibinfo {volume} {66}},\ \bibinfo {pages} {241} (\bibinfo {year} {2015})},\ \Eprint {https://arxiv.org/abs/https://doi.org/10.1146/annurev-physchem-040214-121307} {https://doi.org/10.1146/annurev-physchem-040214-121307} \BibitemShut {NoStop}%
\bibitem [{\citenamefont {Kraus}\ \emph {et~al.}(2015)\citenamefont {Kraus}, \citenamefont {Tamtögl}, \citenamefont {Mayrhofer-Reinhartshuber}, \citenamefont {Apolloner}, \citenamefont {Gösweiner}, \citenamefont {Miret-Artés},\ and\ \citenamefont {Ernst}}]{bismuthhas}%
  \BibitemOpen
  \bibfield  {author} {\bibinfo {author} {\bibfnamefont {P.}~\bibnamefont {Kraus}}, \bibinfo {author} {\bibfnamefont {A.}~\bibnamefont {Tamtögl}}, \bibinfo {author} {\bibfnamefont {M.}~\bibnamefont {Mayrhofer-Reinhartshuber}}, \bibinfo {author} {\bibfnamefont {F.}~\bibnamefont {Apolloner}}, \bibinfo {author} {\bibfnamefont {C.}~\bibnamefont {Gösweiner}}, \bibinfo {author} {\bibfnamefont {S.}~\bibnamefont {Miret-Artés}},\ and\ \bibinfo {author} {\bibfnamefont {W.}~\bibnamefont {Ernst}},\ }\bibfield  {title} {\bibinfo {title} {Surface structure of {Bi(111)} from helium atom scattering measurements - inelastic close-coupling formalism},\ }\href {https://doi.org/10.1021/acs.jpcc.5b05010} {\bibfield  {journal} {\bibinfo  {journal} {J. Phys. Chem. C}\ }\textbf {\bibinfo {volume} {119}},\ \bibinfo {pages} {17235} (\bibinfo {year} {2015})},\ \Eprint {https://arxiv.org/abs/https://doi.org/10.1021/acs.jpcc.5b05010} {https://doi.org/10.1021/acs.jpcc.5b05010} \BibitemShut {NoStop}%
\bibitem [{\citenamefont {Liu}\ \emph {et~al.}(2024{\natexlab{a}})\citenamefont {Liu}, \citenamefont {Allison}, \citenamefont {Peng}, \citenamefont {Avidor}, \citenamefont {Monserrat},\ and\ \citenamefont {Jardine}}]{liu_distinguishing_2024}%
  \BibitemOpen
  \bibfield  {author} {\bibinfo {author} {\bibfnamefont {B.}~\bibnamefont {Liu}}, \bibinfo {author} {\bibfnamefont {W.}~\bibnamefont {Allison}}, \bibinfo {author} {\bibfnamefont {B.}~\bibnamefont {Peng}}, \bibinfo {author} {\bibfnamefont {N.}~\bibnamefont {Avidor}}, \bibinfo {author} {\bibfnamefont {B.}~\bibnamefont {Monserrat}},\ and\ \bibinfo {author} {\bibfnamefont {A.~P.}\ \bibnamefont {Jardine}},\ }\bibfield  {title} {\bibinfo {title} {Distinguishing {Quasiparticle}-{Phonon} {Interactions} by {Ultrahigh}-{Resolution} {Lifetime} {Measurements}},\ }\href {https://doi.org/10.1103/PhysRevLett.132.176202} {\bibfield  {journal} {\bibinfo  {journal} {Physical Review Letters}\ }\textbf {\bibinfo {volume} {132}},\ \bibinfo {pages} {176202} (\bibinfo {year} {2024}{\natexlab{a}})},\ \bibinfo {note} {publisher: American Physical Society}\BibitemShut {NoStop}%
\bibitem [{\citenamefont {Liu}\ \emph {et~al.}(2024{\natexlab{b}})\citenamefont {Liu}, \citenamefont {Kelsall}, \citenamefont {Ward},\ and\ \citenamefont {Jardine}}]{liu_experimental_2024}%
  \BibitemOpen
  \bibfield  {author} {\bibinfo {author} {\bibfnamefont {B.}~\bibnamefont {Liu}}, \bibinfo {author} {\bibfnamefont {J.}~\bibnamefont {Kelsall}}, \bibinfo {author} {\bibfnamefont {D.~J.}\ \bibnamefont {Ward}},\ and\ \bibinfo {author} {\bibfnamefont {A.~P.}\ \bibnamefont {Jardine}},\ }\bibfield  {title} {\bibinfo {title} {Experimental {Characterization} of {Defect}-{Induced} {Phonon} {Lifetime} {Shortening}},\ }\href {https://doi.org/10.1103/PhysRevLett.132.056202} {\bibfield  {journal} {\bibinfo  {journal} {Physical Review Letters}\ }\textbf {\bibinfo {volume} {132}},\ \bibinfo {pages} {056202} (\bibinfo {year} {2024}{\natexlab{b}})},\ \bibinfo {note} {publisher: American Physical Society}\BibitemShut {NoStop}%
\bibitem [{\citenamefont {Scoles}(1988)}]{scoles_beam_methods}%
  \BibitemOpen
  \bibfield  {author} {\bibinfo {author} {\bibfnamefont {G.}~\bibnamefont {Scoles}},\ }\href@noop {} {\emph {\bibinfo {title} {Atomic and molecular beam methods / Vol.1.}}}\ (\bibinfo  {publisher} {Oxford University Press},\ \bibinfo {year} {1988})\BibitemShut {NoStop}%
\bibitem [{\citenamefont {Morse}(1996)}]{generalfreejet}%
  \BibitemOpen
  \bibfield  {author} {\bibinfo {author} {\bibfnamefont {M.~D.}\ \bibnamefont {Morse}},\ }\bibfield  {title} {\bibinfo {title} {2 - supersonic beam sources},\ }\href@noop {} {\bibfield  {journal} {\bibinfo  {journal} {Experimental Methods in The Physical Sciences}\ }\textbf {\bibinfo {volume} {29}},\ \bibinfo {pages} {21} (\bibinfo {year} {1996})}\BibitemShut {NoStop}%
\bibitem [{\citenamefont {Campargue}(1984)}]{campargue}%
  \BibitemOpen
  \bibfield  {author} {\bibinfo {author} {\bibfnamefont {R.}~\bibnamefont {Campargue}},\ }\bibfield  {title} {\bibinfo {title} {Progress in overexpanded supersonic jets and skimmed molecular beams in free-jet zones of silence},\ }\href {https://doi.org/10.1021/j150664a004} {\bibfield  {journal} {\bibinfo  {journal} {J. Phys. Chem.}\ }\textbf {\bibinfo {volume} {88}},\ \bibinfo {pages} {4466} (\bibinfo {year} {1984})},\ \Eprint {https://arxiv.org/abs/https://doi.org/10.1021/j150664a004} {https://doi.org/10.1021/j150664a004} \BibitemShut {NoStop}%
\bibitem [{\citenamefont {Khalil}\ and\ \citenamefont {Miller}(2004)}]{Khalil2004}%
  \BibitemOpen
  \bibfield  {author} {\bibinfo {author} {\bibfnamefont {I.}~\bibnamefont {Khalil}}\ and\ \bibinfo {author} {\bibfnamefont {D.~R.}\ \bibnamefont {Miller}},\ }\bibfield  {title} {\bibinfo {title} {The structure of supercritical fluid free‐jet expansions},\ }\href {https://doi.org/10.1002/aic.10285} {\bibfield  {journal} {\bibinfo  {journal} {AIChE Journal}\ }\textbf {\bibinfo {volume} {50}},\ \bibinfo {pages} {2697–2704} (\bibinfo {year} {2004})}\BibitemShut {NoStop}%
\bibitem [{\citenamefont {Jugroot}\ \emph {et~al.}(2004)\citenamefont {Jugroot}, \citenamefont {Groth}, \citenamefont {Thomson}, \citenamefont {Baranov},\ and\ \citenamefont {Collings}}]{Jugroot2004}%
  \BibitemOpen
  \bibfield  {author} {\bibinfo {author} {\bibfnamefont {M.}~\bibnamefont {Jugroot}}, \bibinfo {author} {\bibfnamefont {C.~P.~T.}\ \bibnamefont {Groth}}, \bibinfo {author} {\bibfnamefont {B.~A.}\ \bibnamefont {Thomson}}, \bibinfo {author} {\bibfnamefont {V.}~\bibnamefont {Baranov}},\ and\ \bibinfo {author} {\bibfnamefont {B.~A.}\ \bibnamefont {Collings}},\ }\bibfield  {title} {\bibinfo {title} {Numerical investigation of interface region flows in mass spectrometers: neutral gas transport},\ }\href {https://doi.org/10.1088/0022-3727/37/8/019} {\bibfield  {journal} {\bibinfo  {journal} {Journal of Physics D: Applied Physics}\ }\textbf {\bibinfo {volume} {37}},\ \bibinfo {pages} {1289–1300} (\bibinfo {year} {2004})}\BibitemShut {NoStop}%
\bibitem [{\citenamefont {Bird}(1976)}]{bird}%
  \BibitemOpen
  \bibfield  {author} {\bibinfo {author} {\bibfnamefont {G.~A.}\ \bibnamefont {Bird}},\ }\bibfield  {title} {\bibinfo {title} {Transition regime behavior of supersonic beam skimmers},\ }\href {https://doi.org/10.1063/1.861351} {\bibfield  {journal} {\bibinfo  {journal} {Phys. Fluids}\ }\textbf {\bibinfo {volume} {19}},\ \bibinfo {pages} {1486} (\bibinfo {year} {1976})},\ \Eprint {https://arxiv.org/abs/https://aip.scitation.org/doi/pdf/10.1063/1.861351} {https://aip.scitation.org/doi/pdf/10.1063/1.861351} \BibitemShut {NoStop}%
\bibitem [{\citenamefont {Eder}\ \emph {et~al.}(2018)\citenamefont {Eder}, \citenamefont {Palau}, \citenamefont {Kaltenbacher}, \citenamefont {Bracco},\ and\ \citenamefont {Holst}}]{microskimmersr}%
  \BibitemOpen
  \bibfield  {author} {\bibinfo {author} {\bibfnamefont {S.~D.}\ \bibnamefont {Eder}}, \bibinfo {author} {\bibfnamefont {A.~S.}\ \bibnamefont {Palau}}, \bibinfo {author} {\bibfnamefont {T.}~\bibnamefont {Kaltenbacher}}, \bibinfo {author} {\bibfnamefont {G.}~\bibnamefont {Bracco}},\ and\ \bibinfo {author} {\bibfnamefont {B.}~\bibnamefont {Holst}},\ }\bibfield  {title} {\bibinfo {title} {Velocity distributions in microskimmer supersonic expansion helium beams: High precision measurements and modeling},\ }\href {https://doi.org/10.1063/1.5044203} {\bibfield  {journal} {\bibinfo  {journal} {Rev. of Sci. Instrum.}\ }\textbf {\bibinfo {volume} {89}},\ \bibinfo {pages} {113301} (\bibinfo {year} {2018})},\ \Eprint {https://arxiv.org/abs/https://doi.org/10.1063/1.5044203} {https://doi.org/10.1063/1.5044203} \BibitemShut {NoStop}%
\bibitem [{\citenamefont {Palau}\ \emph {et~al.}(2018{\natexlab{a}})\citenamefont {Palau}, \citenamefont {Eder}, \citenamefont {Andersen}, \citenamefont {Ravn}, \citenamefont {Bracco},\ and\ \citenamefont {Holst}}]{palauintensity}%
  \BibitemOpen
  \bibfield  {author} {\bibinfo {author} {\bibfnamefont {A.~S.}\ \bibnamefont {Palau}}, \bibinfo {author} {\bibfnamefont {S.~D.}\ \bibnamefont {Eder}}, \bibinfo {author} {\bibfnamefont {T.}~\bibnamefont {Andersen}}, \bibinfo {author} {\bibfnamefont {A.~K.}\ \bibnamefont {Ravn}}, \bibinfo {author} {\bibfnamefont {G.}~\bibnamefont {Bracco}},\ and\ \bibinfo {author} {\bibfnamefont {B.}~\bibnamefont {Holst}},\ }\bibfield  {title} {\bibinfo {title} {Center-line intensity of a supersonic helium beam},\ }\href {https://doi.org/10.1103/PhysRevA.98.063611} {\bibfield  {journal} {\bibinfo  {journal} {Phys. Rev. A}\ }\textbf {\bibinfo {volume} {98}},\ \bibinfo {pages} {063611} (\bibinfo {year} {2018}{\natexlab{a}})}\BibitemShut {NoStop}%
\bibitem [{\citenamefont {Yamashita}\ and\ \citenamefont {Fenn}(1984)}]{Yamashita1984}%
  \BibitemOpen
  \bibfield  {author} {\bibinfo {author} {\bibfnamefont {M.}~\bibnamefont {Yamashita}}\ and\ \bibinfo {author} {\bibfnamefont {J.~B.}\ \bibnamefont {Fenn}},\ }\bibfield  {title} {\bibinfo {title} {Electrospray ion source. another variation on the free-jet theme},\ }\href {https://doi.org/10.1021/j150664a002} {\bibfield  {journal} {\bibinfo  {journal} {The Journal of Physical Chemistry}\ }\textbf {\bibinfo {volume} {88}},\ \bibinfo {pages} {4451–4459} (\bibinfo {year} {1984})}\BibitemShut {NoStop}%
\bibitem [{\citenamefont {Fenn}(2000)}]{Fenn2000}%
  \BibitemOpen
  \bibfield  {author} {\bibinfo {author} {\bibfnamefont {J.~B.}\ \bibnamefont {Fenn}},\ }\bibfield  {title} {\bibinfo {title} {Mass spectrometric implications of high-pressure ion sources},\ }\href {https://doi.org/10.1016/s1387-3806(00)00328-6} {\bibfield  {journal} {\bibinfo  {journal} {International Journal of Mass Spectrometry}\ }\textbf {\bibinfo {volume} {200}},\ \bibinfo {pages} {459–478} (\bibinfo {year} {2000})}\BibitemShut {NoStop}%
\bibitem [{\citenamefont {Beijerinck}\ \emph {et~al.}(1985)\citenamefont {Beijerinck}, \citenamefont {van Gerwen}, \citenamefont {Kerstel}, \citenamefont {Martens}, \citenamefont {van Vliembergen}, \citenamefont {r.~Th.~Smits},\ and\ \citenamefont {Kaashoek}}]{oldcamparguemodelling}%
  \BibitemOpen
  \bibfield  {author} {\bibinfo {author} {\bibfnamefont {H.~C.~W.}\ \bibnamefont {Beijerinck}}, \bibinfo {author} {\bibfnamefont {R.~J.~F.}\ \bibnamefont {van Gerwen}}, \bibinfo {author} {\bibfnamefont {E.~R.~T.}\ \bibnamefont {Kerstel}}, \bibinfo {author} {\bibfnamefont {J.~F.~M.}\ \bibnamefont {Martens}}, \bibinfo {author} {\bibfnamefont {E.~J.~W.}\ \bibnamefont {van Vliembergen}}, \bibinfo {author} {\bibfnamefont {M.}~\bibnamefont {r.~Th.~Smits}},\ and\ \bibinfo {author} {\bibfnamefont {G.~H.}\ \bibnamefont {Kaashoek}},\ }\bibfield  {title} {\bibinfo {title} {Campargue-type supersonic beam sources: Absolute intensities, skimmer transmission and scaling laws for mono-atomic gases {He}, {Ne} and {Ar}},\ }\href {https://doi.org/https://doi.org/10.1016/0301-0104(85)80201-9} {\bibfield  {journal} {\bibinfo  {journal} {Chem. Phys.}\ }\textbf {\bibinfo {volume} {96}},\ \bibinfo {pages} {153} (\bibinfo {year} {1985})}\BibitemShut {NoStop}%
\bibitem [{\citenamefont {Segev}\ \emph {et~al.}(2017)\citenamefont {Segev}, \citenamefont {Bibelnik}, \citenamefont {Akerman}, \citenamefont {Shagam}, \citenamefont {Luski}, \citenamefont {Karpov}, \citenamefont {Narevicius},\ and\ \citenamefont {Narevicius}}]{dsmcbrightening}%
  \BibitemOpen
  \bibfield  {author} {\bibinfo {author} {\bibfnamefont {Y.}~\bibnamefont {Segev}}, \bibinfo {author} {\bibfnamefont {N.}~\bibnamefont {Bibelnik}}, \bibinfo {author} {\bibfnamefont {N.}~\bibnamefont {Akerman}}, \bibinfo {author} {\bibfnamefont {Y.}~\bibnamefont {Shagam}}, \bibinfo {author} {\bibfnamefont {A.}~\bibnamefont {Luski}}, \bibinfo {author} {\bibfnamefont {M.}~\bibnamefont {Karpov}}, \bibinfo {author} {\bibfnamefont {J.}~\bibnamefont {Narevicius}},\ and\ \bibinfo {author} {\bibfnamefont {E.}~\bibnamefont {Narevicius}},\ }\bibfield  {title} {\bibinfo {title} {Molecular beam brightening by shock-wave suppression},\ }\href {https://doi.org/10.1126/sciadv.1602258} {\bibfield  {journal} {\bibinfo  {journal} {Sci. Adv.}\ }\textbf {\bibinfo {volume} {3}},\ \bibinfo {pages} {1602258} (\bibinfo {year} {2017})},\ \Eprint {https://arxiv.org/abs/https://www.science.org/doi/pdf/10.1126/sciadv.1602258} {https://www.science.org/doi/pdf/10.1126/sciadv.1602258} \BibitemShut {NoStop}%
\bibitem [{\citenamefont {Hedgeland}\ \emph {et~al.}(2005)\citenamefont {Hedgeland}, \citenamefont {Jardine}, \citenamefont {Allison},\ and\ \citenamefont {Ellis}}]{anomalous}%
  \BibitemOpen
  \bibfield  {author} {\bibinfo {author} {\bibfnamefont {H.}~\bibnamefont {Hedgeland}}, \bibinfo {author} {\bibfnamefont {A.~P.}\ \bibnamefont {Jardine}}, \bibinfo {author} {\bibfnamefont {W.}~\bibnamefont {Allison}},\ and\ \bibinfo {author} {\bibfnamefont {J.}~\bibnamefont {Ellis}},\ }\bibfield  {title} {\bibinfo {title} {Anomalous attenuation at low temperatures in high-intensity helium beam sources},\ }\href {https://doi.org/10.1063/1.2149008} {\bibfield  {journal} {\bibinfo  {journal} {Rev. Sci. Intr.}\ }\textbf {\bibinfo {volume} {76}},\ \bibinfo {pages} {123111} (\bibinfo {year} {2005})}\BibitemShut {NoStop}%
\bibitem [{\citenamefont {Verheijen}\ \emph {et~al.}(1984)\citenamefont {Verheijen}, \citenamefont {Beijerinck}, \citenamefont {Renes},\ and\ \citenamefont {Verster}}]{oldmodelling}%
  \BibitemOpen
  \bibfield  {author} {\bibinfo {author} {\bibfnamefont {M.~J.}\ \bibnamefont {Verheijen}}, \bibinfo {author} {\bibfnamefont {H.~C.~W.}\ \bibnamefont {Beijerinck}}, \bibinfo {author} {\bibfnamefont {W.~A.}\ \bibnamefont {Renes}},\ and\ \bibinfo {author} {\bibfnamefont {N.~F.}\ \bibnamefont {Verster}},\ }\bibfield  {title} {\bibinfo {title} {{A quantitative description of skimmer interaction in supersonic secondary beams: Calibration of absolute intensities}},\ }\href {https://doi.org/10.1016/S0301-0104(84)85173-3} {\bibfield  {journal} {\bibinfo  {journal} {Chem. Phys.}\ }\textbf {\bibinfo {volume} {85}},\ \bibinfo {pages} {63} (\bibinfo {year} {1984})}\BibitemShut {NoStop}%
\bibitem [{\citenamefont {Mukherjee}\ \emph {et~al.}(2000)\citenamefont {Mukherjee}, \citenamefont {Gantayet},\ and\ \citenamefont {Ahmad}}]{somedsmcsimulations}%
  \BibitemOpen
  \bibfield  {author} {\bibinfo {author} {\bibfnamefont {J.}~\bibnamefont {Mukherjee}}, \bibinfo {author} {\bibfnamefont {L.~M.}\ \bibnamefont {Gantayet}},\ and\ \bibinfo {author} {\bibfnamefont {S.~A.}\ \bibnamefont {Ahmad}},\ }\bibfield  {title} {\bibinfo {title} {Free jet expansion of atomic beam: simulation studies of some parameters},\ }\href {https://doi.org/10.1088/0022-3727/33/11/319} {\bibfield  {journal} {\bibinfo  {journal} {J. Phys. D: Appl. Phys.}\ }\textbf {\bibinfo {volume} {33}},\ \bibinfo {pages} {1386} (\bibinfo {year} {2000})}\BibitemShut {NoStop}%
\bibitem [{\citenamefont {Montero}(2017)}]{hybridfluids}%
  \BibitemOpen
  \bibfield  {author} {\bibinfo {author} {\bibfnamefont {S.}~\bibnamefont {Montero}},\ }\bibfield  {title} {\bibinfo {title} {Molecular description of steady supersonic free jets},\ }\href {https://doi.org/10.1063/1.5001250} {\bibfield  {journal} {\bibinfo  {journal} {Phys. Fluids}\ }\textbf {\bibinfo {volume} {29}},\ \bibinfo {pages} {096101} (\bibinfo {year} {2017})},\ \Eprint {https://arxiv.org/abs/https://doi.org/10.1063/1.5001250} {https://doi.org/10.1063/1.5001250} \BibitemShut {NoStop}%
\bibitem [{\citenamefont {Gould}\ and\ \citenamefont {Bučko}(2016)}]{c6valuescomput}%
  \BibitemOpen
  \bibfield  {author} {\bibinfo {author} {\bibfnamefont {T.}~\bibnamefont {Gould}}\ and\ \bibinfo {author} {\bibfnamefont {T.}~\bibnamefont {Bučko}},\ }\bibfield  {title} {\bibinfo {title} {C6 coefficients and dipole polarizabilities for all atoms and many ions in rows 1–6 of the periodic table},\ }\href {https://doi.org/10.1021/acs.jctc.6b00361} {\bibfield  {journal} {\bibinfo  {journal} {J. Chem. Theory. Comput.}\ }\textbf {\bibinfo {volume} {12}},\ \bibinfo {pages} {3603} (\bibinfo {year} {2016})},\ \Eprint {https://arxiv.org/abs/https://doi.org/10.1021/acs.jctc.6b00361} {https://doi.org/10.1021/acs.jctc.6b00361} \BibitemShut {NoStop}%
\bibitem [{\citenamefont {Beijerinck}\ and\ \citenamefont {Verster}(1981)}]{modellingsupersonicdistribution}%
  \BibitemOpen
  \bibfield  {author} {\bibinfo {author} {\bibfnamefont {H.~C.~W.}\ \bibnamefont {Beijerinck}}\ and\ \bibinfo {author} {\bibfnamefont {N.~F.}\ \bibnamefont {Verster}},\ }\bibfield  {title} {\bibinfo {title} {Absolute intensities and perpendicular temperatures of supersonic beams of polyatomic gases},\ }\href {https://doi.org/https://doi.org/10.1016/0378-4363(81)90112-1} {\bibfield  {journal} {\bibinfo  {journal} {Physica B+C}\ }\textbf {\bibinfo {volume} {111}},\ \bibinfo {pages} {327} (\bibinfo {year} {1981})}\BibitemShut {NoStop}%
\bibitem [{\citenamefont {Lambrick}\ \emph {et~al.}(2018)\citenamefont {Lambrick}, \citenamefont {Bergin}, \citenamefont {Jardine},\ and\ \citenamefont {Ward}}]{raytracing}%
  \BibitemOpen
  \bibfield  {author} {\bibinfo {author} {\bibfnamefont {S.~M.}\ \bibnamefont {Lambrick}}, \bibinfo {author} {\bibfnamefont {M.}~\bibnamefont {Bergin}}, \bibinfo {author} {\bibfnamefont {A.~P.}\ \bibnamefont {Jardine}},\ and\ \bibinfo {author} {\bibfnamefont {D.~J.}\ \bibnamefont {Ward}},\ }\bibfield  {title} {\bibinfo {title} {A ray tracing method for predicting contrast in neutral atom beam imaging},\ }\href {https://doi.org/https://doi.org/10.1016/j.micron.2018.06.014} {\bibfield  {journal} {\bibinfo  {journal} {Micron}\ }\textbf {\bibinfo {volume} {113}},\ \bibinfo {pages} {61} (\bibinfo {year} {2018})}\BibitemShut {NoStop}%
\bibitem [{\citenamefont {Fahy}\ \emph {et~al.}(2018)\citenamefont {Fahy}, \citenamefont {Eder}, \citenamefont {Barr}, \citenamefont {Martens}, \citenamefont {Myles},\ and\ \citenamefont {Dastoor}}]{image}%
  \BibitemOpen
  \bibfield  {author} {\bibinfo {author} {\bibfnamefont {A.}~\bibnamefont {Fahy}}, \bibinfo {author} {\bibfnamefont {S.~D.}\ \bibnamefont {Eder}}, \bibinfo {author} {\bibfnamefont {M.}~\bibnamefont {Barr}}, \bibinfo {author} {\bibfnamefont {J.}~\bibnamefont {Martens}}, \bibinfo {author} {\bibfnamefont {T.~A.}\ \bibnamefont {Myles}},\ and\ \bibinfo {author} {\bibfnamefont {P.~C.}\ \bibnamefont {Dastoor}},\ }\bibfield  {title} {\bibinfo {title} {Image formation in the scanning helium microscope},\ }\href {https://doi.org/https://doi.org/10.1016/j.ultramic.2018.05.004} {\bibfield  {journal} {\bibinfo  {journal} {Ultramicroscopy}\ }\textbf {\bibinfo {volume} {192}},\ \bibinfo {pages} {7} (\bibinfo {year} {2018})}\BibitemShut {NoStop}%
\bibitem [{\citenamefont {Eder}\ \emph {et~al.}(2023)\citenamefont {Eder}, \citenamefont {Fahy}, \citenamefont {Barr}, \citenamefont {Manson}, \citenamefont {Holst},\ and\ \citenamefont {Dastoor}}]{Eder2023}%
  \BibitemOpen
  \bibfield  {author} {\bibinfo {author} {\bibfnamefont {S.~D.}\ \bibnamefont {Eder}}, \bibinfo {author} {\bibfnamefont {A.}~\bibnamefont {Fahy}}, \bibinfo {author} {\bibfnamefont {M.~G.}\ \bibnamefont {Barr}}, \bibinfo {author} {\bibfnamefont {J.~R.}\ \bibnamefont {Manson}}, \bibinfo {author} {\bibfnamefont {B.}~\bibnamefont {Holst}},\ and\ \bibinfo {author} {\bibfnamefont {P.~C.}\ \bibnamefont {Dastoor}},\ }\bibfield  {title} {\bibinfo {title} {Sub-resolution contrast in neutral helium microscopy through facet scattering for quantitative imaging of nanoscale topographies on macroscopic surfaces},\ }\href {https://doi.org/10.1038/s41467-023-36578-x} {\bibfield  {journal} {\bibinfo  {journal} {Nature Communications}\ }\textbf {\bibinfo {volume} {14}},\ \bibinfo {pages} {36578} (\bibinfo {year} {2023})}\BibitemShut {NoStop}%
\bibitem [{\citenamefont {Lambrick}\ \emph {et~al.}(2022{\natexlab{b}})\citenamefont {Lambrick}, \citenamefont {Bergin}, \citenamefont {Ward}, \citenamefont {Barr}, \citenamefont {Fahy}, \citenamefont {Myles}, \citenamefont {Radić}, \citenamefont {Dastoor}, \citenamefont {Ellis},\ and\ \citenamefont {Jardine}}]{samcosinescattering}%
  \BibitemOpen
  \bibfield  {author} {\bibinfo {author} {\bibfnamefont {S.~M.}\ \bibnamefont {Lambrick}}, \bibinfo {author} {\bibfnamefont {M.}~\bibnamefont {Bergin}}, \bibinfo {author} {\bibfnamefont {D.~J.}\ \bibnamefont {Ward}}, \bibinfo {author} {\bibfnamefont {M.}~\bibnamefont {Barr}}, \bibinfo {author} {\bibfnamefont {A.}~\bibnamefont {Fahy}}, \bibinfo {author} {\bibfnamefont {T.}~\bibnamefont {Myles}}, \bibinfo {author} {\bibfnamefont {A.}~\bibnamefont {Radić}}, \bibinfo {author} {\bibfnamefont {P.~C.}\ \bibnamefont {Dastoor}}, \bibinfo {author} {\bibfnamefont {J.}~\bibnamefont {Ellis}},\ and\ \bibinfo {author} {\bibfnamefont {A.~P.}\ \bibnamefont {Jardine}},\ }\bibfield  {title} {\bibinfo {title} {Observation of diffuse scattering in scanning helium microscopy},\ }\href {https://doi.org/10.1039/D2CP01951E} {\bibfield  {journal} {\bibinfo  {journal} {Phys. Chem. Chem. Phys.}\ }\textbf {\bibinfo {volume} {24}},\ \bibinfo {pages} {26539} (\bibinfo {year} {2022}{\natexlab{b}})}\BibitemShut {NoStop}%
\bibitem [{\citenamefont {Saxena}\ and\ \citenamefont {Joshi}(1982)}]{thermalaccomodation}%
  \BibitemOpen
  \bibfield  {author} {\bibinfo {author} {\bibfnamefont {S.~C.}\ \bibnamefont {Saxena}}\ and\ \bibinfo {author} {\bibfnamefont {R.~K.}\ \bibnamefont {Joshi}},\ }\href@noop {} {\emph {\bibinfo {title} {Thermal Accommodation and Adsorption Coefficients of Gases}}}\ (\bibinfo  {publisher} {McGraw-Hill},\ \bibinfo {year} {1982})\ pp.\ \bibinfo {pages} {199, 262}\BibitemShut {NoStop}%
\bibitem [{\citenamefont {Yasumoto}(1987)}]{lambdadefinition}%
  \BibitemOpen
  \bibfield  {author} {\bibinfo {author} {\bibfnamefont {I.}~\bibnamefont {Yasumoto}},\ }\bibfield  {title} {\bibinfo {title} {Accommodation coefficients of helium, neon, argon, hydrogen, and deuterium on graphitized carbon},\ }\href {https://doi.org/10.1021/j100300a019} {\bibfield  {journal} {\bibinfo  {journal} {J. Phys. Chem}\ }\textbf {\bibinfo {volume} {91}},\ \bibinfo {pages} {4298} (\bibinfo {year} {1987})},\ \Eprint {https://arxiv.org/abs/https://doi.org/10.1021/j100300a019} {https://doi.org/10.1021/j100300a019} \BibitemShut {NoStop}%
\bibitem [{\citenamefont {Lechner}\ \emph {et~al.}(2013)\citenamefont {Lechner}, \citenamefont {Hedgeland}, \citenamefont {Allison}, \citenamefont {Ellis},\ and\ \citenamefont {Jardine}}]{thesource}%
  \BibitemOpen
  \bibfield  {author} {\bibinfo {author} {\bibfnamefont {B.~A.~J.}\ \bibnamefont {Lechner}}, \bibinfo {author} {\bibfnamefont {H.}~\bibnamefont {Hedgeland}}, \bibinfo {author} {\bibfnamefont {W.}~\bibnamefont {Allison}}, \bibinfo {author} {\bibfnamefont {J.}~\bibnamefont {Ellis}},\ and\ \bibinfo {author} {\bibfnamefont {A.~P.}\ \bibnamefont {Jardine}},\ }\bibfield  {title} {\bibinfo {title} {Note: A new design for a low-temperature high-intensity helium beam source},\ }\href {https://doi.org/10.1063/1.4791929} {\bibfield  {journal} {\bibinfo  {journal} {Rev. Sci. Instrum.}\ }\textbf {\bibinfo {volume} {84}},\ \bibinfo {pages} {026105} (\bibinfo {year} {2013})},\ \Eprint {https://arxiv.org/abs/https://doi.org/10.1063/1.4791929} {https://doi.org/10.1063/1.4791929} \BibitemShut {NoStop}%
\bibitem [{\citenamefont {Lechner}(2012)}]{bajl}%
  \BibitemOpen
  \bibfield  {author} {\bibinfo {author} {\bibfnamefont {B.~A.~J.}\ \bibnamefont {Lechner}},\ }\emph {\bibinfo {title} {Studying complex surface dynamical systems using helium-3 spin-echo spectroscopy}},\ \href@noop {} {Ph.D. thesis},\ \bibinfo  {school} {University of Cambridge} (\bibinfo {year} {2012})\BibitemShut {NoStop}%
\bibitem [{\citenamefont {Young}(1975)}]{Machdisctheory}%
  \BibitemOpen
  \bibfield  {author} {\bibinfo {author} {\bibfnamefont {W.~S.}\ \bibnamefont {Young}},\ }\bibfield  {title} {\bibinfo {title} {Derivation of the free-jet mach-disk location using the entropy-balance principle},\ }\href {https://doi.org/10.1063/1.861039} {\bibfield  {journal} {\bibinfo  {journal} {Phys. Fluids}\ }\textbf {\bibinfo {volume} {18}},\ \bibinfo {pages} {1421} (\bibinfo {year} {1975})},\ \Eprint {https://arxiv.org/abs/https://aip.scitation.org/doi/pdf/10.1063/1.861039} {https://aip.scitation.org/doi/pdf/10.1063/1.861039} \BibitemShut {NoStop}%
\bibitem [{\citenamefont {Jardine}(2001)}]{andy}%
  \BibitemOpen
  \bibfield  {author} {\bibinfo {author} {\bibfnamefont {A.}~\bibnamefont {Jardine}},\ }\emph {\bibinfo {title} {Quasi-elastic Helium Atom Scattering: Interpretation and Instrumentation}},\ \href@noop {} {Ph.D. thesis},\ \bibinfo  {school} {University of Cambridge} (\bibinfo {year} {2001})\BibitemShut {NoStop}%
\bibitem [{\citenamefont {Chisnall}(2012)}]{chisnall}%
  \BibitemOpen
  \bibfield  {author} {\bibinfo {author} {\bibfnamefont {D.}~\bibnamefont {Chisnall}},\ }\emph {\bibinfo {title} {A High Sensitivity Detector for Helium Atom Scattering}},\ \href@noop {} {Ph.D. thesis},\ \bibinfo  {school} {University of Cambridge} (\bibinfo {year} {2012})\BibitemShut {NoStop}%
\bibitem [{\citenamefont {Radic}\ \emph {et~al.}(2023)\citenamefont {Radic}, \citenamefont {Lambrick}, \citenamefont {Rhodes},\ and\ \citenamefont {Ward}}]{Radic_plastics_2023}%
  \BibitemOpen
  \bibfield  {author} {\bibinfo {author} {\bibfnamefont {A.}~\bibnamefont {Radic}}, \bibinfo {author} {\bibfnamefont {S.}~\bibnamefont {Lambrick}}, \bibinfo {author} {\bibfnamefont {S.}~\bibnamefont {Rhodes}},\ and\ \bibinfo {author} {\bibfnamefont {D.~J.}\ \bibnamefont {Ward}},\ }\bibfield  {title} {\bibinfo {title} {On the application of components manufactured with stereolithographic 3d printing in high vacuum systems},\ }\bibfield  {journal} {\bibinfo  {journal} {Vacuum}\ }\href {https://doi.org/10.21203/rs.3.rs-3681351/v1} {10.21203/rs.3.rs-3681351/v1} (\bibinfo {year} {2023}),\ \bibinfo {note} {under Review}\BibitemShut {NoStop}%
\bibitem [{\citenamefont {Palau}\ \emph {et~al.}(2018{\natexlab{b}})\citenamefont {Palau}, \citenamefont {Eder}, \citenamefont {Andersen}, \citenamefont {Ravn}, \citenamefont {Bracco},\ and\ \citenamefont {Holst}}]{adria}%
  \BibitemOpen
  \bibfield  {author} {\bibinfo {author} {\bibfnamefont {A.~S.}\ \bibnamefont {Palau}}, \bibinfo {author} {\bibfnamefont {S.~D.}\ \bibnamefont {Eder}}, \bibinfo {author} {\bibfnamefont {T.}~\bibnamefont {Andersen}}, \bibinfo {author} {\bibfnamefont {A.~K.}\ \bibnamefont {Ravn}}, \bibinfo {author} {\bibfnamefont {G.}~\bibnamefont {Bracco}},\ and\ \bibinfo {author} {\bibfnamefont {B.}~\bibnamefont {Holst}},\ }\bibfield  {title} {\bibinfo {title} {Center-line intensity of a supersonic helium beam},\ }\href {https://doi.org/10.1103/PhysRevA.98.063611} {\bibfield  {journal} {\bibinfo  {journal} {Phys. Rev. A}\ }\textbf {\bibinfo {volume} {98}},\ \bibinfo {pages} {063611} (\bibinfo {year} {2018}{\natexlab{b}})}\BibitemShut {NoStop}%
\bibitem [{\citenamefont {inc.}(2024{\natexlab{a}})}]{beamDynamicsHomepage}%
  \BibitemOpen
  \bibfield  {author} {\bibinfo {author} {\bibfnamefont {B.~D.}\ \bibnamefont {inc.}},\ }\href {https://www.beamdynamicsinc.com/} {\bibinfo {title} {Beam dynamics home page}} (\bibinfo {year} {2024}{\natexlab{a}})\BibitemShut {NoStop}%
\bibitem [{\citenamefont {inc.}(2024{\natexlab{b}})}]{beamDynamicsSkimmerSpecs}%
  \BibitemOpen
  \bibfield  {author} {\bibinfo {author} {\bibfnamefont {B.~D.}\ \bibnamefont {inc.}},\ }\href {https://www.beamdynamicsinc.com/skimmer-dimensions} {\bibinfo {title} {Skimmer dimensions}} (\bibinfo {year} {2024}{\natexlab{b}})\BibitemShut {NoStop}%
\end{thebibliography}%


\end{document}